\documentclass[graybox]{svmult}
\usepackage{mathptmx}       
\usepackage{helvet}         
\usepackage{courier}        
\usepackage{makeidx}         
\usepackage{graphicx}        
\usepackage{multicol}        
\usepackage[bottom]{footmisc}
\usepackage{amsmath,amssymb,amsfonts}        
\makeindex             
\allowdisplaybreaks

\begin{document}

\title*{On five-dimensional curvature squared supergravity and holography}
\author{Gregory Gold, Peng-Ju Hu, Jessica Hutomo, Saurish Khandelwal, Mehmet Ozkan, Yi Pang, Gabriele Tartaglino-Mazzucchelli}
\institute{Gregory Gold, Gabriele Tartaglino-Mazzucchelli
\at School of Mathematics and Physics, University of Queensland, St Lucia, Brisbane, Queensland 4072, Australia. \\  \email{g.gold@uq.edu.au, g.tartaglino-mazzucchelli@uq.edu.au}
\and Peng-Ju Hu, Yi Pang
\at Center for Joint Quantum Studies and Department of Physics,
School of Science, Tianjin University, Tianjin 300350, China. \\ \email{pengjuhu@tju.edu.cn, pangyi1@tju.edu.cn}
\and Jessica Hutomo
\at INFN, Sezione di Padova, Via Marzolo 8, 35131 Padova, Italy. \\  \email{jessica.hutomo@pd.infn.it}
\and Saurish Khandelwal
\at Institut f\"ur Theoretische Physik Leibniz Universität Hannover Appelstra\ss e 2, 30167 Hannover, Germany. \\  \email{saurish.khandelwal@itp.uni-hannover.de}
\and Mehmet Ozkan
\at Department of Physics, Istanbul Technical University, Maslak 34469 Istanbul, Turkey. \\ \email{ozkanmehm@itu.edu.tr}}

\maketitle
\abstract{
In this work, we report the recent progress in obtaining new curvature-squared invariants in $5D, \mathcal{N}=1$ gauged minimal supergravity. We exhibit the structure of various composite multiplets that are pivotal in the construction. We also present the form of the gauged Riemann-squared and Gauss-Bonnet superinvariants in a dilaton-Weyl multiplet. As a first application of the new curvature squared invariants, we compute their corrections to holographic central charges and the Euclidean action of supersymmetric charged rotating black holes, exhibiting exact matching between the gravity and CFT results.  }


\section{Introduction}

The study of higher-derivative corrections to supergravity theories plays an important role in probing the structure of quantum gravity and string theory. In particular, curvature-squared terms in five-dimensional supergravity emerge as the first stringy and quantum corrections to low-energy effective actions derived from 5D compactifications of string theory. These corrections provide essential tools for exploring quantum aspects of gravity, as well as for refining the ${\rm AdS}_5/{\rm CFT}_4$ correspondence to include leading $\alpha^\prime$ effects, which translate into $1/N$ corrections in the dual conformal field theories (CFTs). Within this framework, a universal sector is described by minimal five-dimensional $\mathcal{N}=1$ gauged supergravity, which is the focus of this work.

Although two-derivative five-dimensional supergravity has been well understood for decades \cite{Cremmer,CN,Howe5Dsugra}, the construction of all independent curvature-squared supergravity invariants in the gauged setting had long remained incomplete. Early progress was made in the ungauged case, where supersymmetric completions of Weyl-squared \cite{HOT} and scalar curvature-squared invariants \cite{Ozkan:2013nwa} were obtained using off-shell conformal supergravity techniques — see \cite{Freedman:2012zz,Lauria:2020rhc,Kuzenko:2022skv,Kuzenko:2022ajd} for reviews on the superconformal tensor calculus for constructing off-shell matter-coupled supergravities, and \cite{Ozkan:2024euj} for its application to higher-derivative supergravities in various dimensions. However, the analyses in \cite{HOT,Ozkan:2013nwa} did not include a supersymmetric extension of the Ricci tensor squared term and were not extended to the gauged case, which is essential for studying supersymmetric (asymptotically) anti-de Sitter (${\rm AdS}_5$) backgrounds and their holographic applications.

These obstacles were recently overcome in a series of works \cite{Gold:2023ymc,Gold:2023dfe,Gold:2023ykx}; see also \cite{Liu:2022sew,Bobev:2022bjm,Cassani:2022lrk} for earlier related developments and applications. In \cite{Gold:2023ymc,Gold:2023dfe,Gold:2023ykx}, we completed the explicit construction of the three independent curvature-squared invariants for gauged $5D$ $\mathcal{N}=1$ supergravity: the supersymmetric completions of the square of the Weyl tensor, the Ricci scalar squared, and a third invariant — often referred to as the ``Log'' multiplet — which contains the Ricci tensor squared term as its leading bosonic contribution. This last invariant had been particularly elusive and required the use of advanced superspace techniques \cite{Butter:2014xxa}, as well as symbolic computer algebra methods \cite{Gold:2024nbw} implemented using the \textit{Cadabra} software \cite{Cadabra-1,Cadabra-2} to fully extract its component structure.

The new constructions mentioned above not only represent progress in formal supersymmetry, but also meet the demand for precision tests of holography beyond leading order. It is widely believed that in the large $N$ expansion, subleading corrections to observables in strongly coupled SCFTs should match the effects of curvature-squared terms in the bulk. An efficient method for evaluating these corrections is thus crucial for comparing gravity and CFT results. In the context of the AdS$_5$/CFT$_4$ correspondence, it was shown in \cite{Ma:2024ynp} that when curvature-squared superinvariants are expressed in terms of the Weyl tensor, corrections to the Euclidean action of charged rotating black holes can be obtained by directly evaluating the four-derivative invariants on the uncorrected solutions. This simplification bypasses the need to solve for the corrected background — a task that is notoriously difficult. From the Euclidean action, other thermodynamic quantities such as the mass, angular momentum, and electric charges can be extracted. Using the AdS/CFT dictionary, it was shown in \cite{Ma:2024ynp,Bobev:2022bjm,Cassani:2022lrk} that the Euclidean action of the BPS black hole matches precisely with the superconformal index of the dual SCFT. Moreover, the linear constraint satisfied by the CFT fugacities, as well as nonlinear constraints among the conserved charges — including their subleading corrections — are reproduced from the gravity side. As a byproduct, the holographic central charges can be directly read off from the boundary logarithmic counterterms. These results provide strong evidence for the robustness of the AdS/CFT correspondence beyond leading order, and emphasize the importance of a complete and consistent higher-derivative supergravity framework.

On one hand, the present paper aims to summarize our contributions to the construction, and recent applications of curvature-squared invariants in five-dimensional minimal gauged supergravity. On the other hand, this paper complements our previous works \cite{Gold:2023ymc,Gold:2023dfe,Gold:2023ykx,Ma:2024ynp} by presenting some new results and analyses. We hope this work serves as a useful reference for both the construction of higher-derivative invariants in supergravity and their applications to problems in quantum gravity and holography.

In section \ref{Curvature-Squared-section}, we begin with a review of the off-shell conformal supergravity formalism and the role of the gauged dilaton Weyl multiplet \cite{Coomans:2012cf} in our recent analysis. We then outline the key ingredients for constructing the three independent gauged curvature-squared invariants of \cite{Gold:2023ymc}, both in component form and via superspace methods. We describe the use of the BF-type supersymmetric action principle, the mechanism of superconformal gauge fixing, and the transition to the gauged off-shell Poincaré theory. In addition to reviewing the results of \cite{Gold:2023ymc}, we also provide the Lagrangians for the off-shell gauged supersymmetric Riemann and Gauss-Bonnet combinations.

In the second part of the paper, we explore applications of these constructions to holography. In section \ref{Holography-section-1}, we study the on-shell structure of the curvature-squared corrections by applying appropriate field redefinitions, which substantially simplify the theory and allow for the extraction of central charges in the dual SCFT. The basis chosen for this on-shell theory differs slightly from that of \cite{Gold:2023ymc} and earlier works \cite{Liu:2022sew,Bobev:2022bjm,Cassani:2022lrk}, as we adopt an on-shell Weyl-squared invariant rather than the Gauss-Bonnet combination as the representative four-derivative term surviving field redefinitions. In section \ref{Holography-section-2}, we review an efficient method for computing curvature-squared corrections to the thermodynamics of charged rotating AdS$_5$ black holes \cite{Ma:2024ynp}. We conclude in section \ref{Conclusion-section} with a brief discussion of open problems and future directions.

\section{Curvature squared invariants}
\label{Curvature-Squared-section}

In this section we review the key results recently obtained in \cite{Gold:2023ymc} pertaining to the construction of gauged curvature-squared supergravity invariants in five dimensions based on the off-shell dilaton Weyl multiplet. For the results in components, our notation and conventions correspond to that of \cite{Bergshoeff:2001hc}. We denote the spacetime indices by $\mu, \nu, \cdots$, Lorentz indices by $a, b, \cdots$, SU(2)  indices by $i, j, \cdots$, and spinor indices by $\alpha, \beta, \cdots$.

\subsection{Standard Weyl}
The standard Weyl multiplet of 5D ${\cal N}=1$ conformal supergravity \cite{Bergshoeff:2001hc} is comprised of a set of independent gauge fields: the vielbein $e_\mu{}^a$,	the gravitino $\psi_{\mu}{}_{\alpha}^i$,	the SU(2) gauge fields $V_\mu{}^{ij}$, and a dilatation gauge field $b_\mu$. The other gauge fields associated with the remaining symmetries, including the spin connection  $\omega_\mu{}^{a b}$, the $S$-supersymmetry connection	$\phi_{\mu}{}_{\alpha}^i$, and the special conformal connection	${f}_\mu{}^a$, are composite fields, i.e., they are determined in terms of the other fields by imposing certain curvature constraints. The standard Weyl multiplet also contains a set of matter fields: a real antisymmetric	tensor $T_{a b}$, a fermion $\chi_\alpha^i$, and a real scalar $D$. A more detailed discussion of the superconformal transformations of the various fields can be found, e.g., in \cite{Bergshoeff:2001hc,Coomans:2012cf}. 

\subsection{Gauged Dilaton-Weyl}
Let us consider a variant multiplet of conformal supergravity, known as the gauged dilaton Weyl multiplet \cite{Coomans:2012cf}.\footnote{Note that the definition of a dilaton Weyl multiplet is not unique. For  instance, in five dimensions a (hyper) dilaton Weyl multiplet based on an  on-shell hypermultiplet in the background of the standard Weyl multiplet was constructed in \cite{Hutomo:2022hdi}, which might also be used for  alternative gaugings. In this paper we work with the gauged (vector) dilaton  Weyl multiplet of \cite{Coomans:2012cf}, see also \cite{Butter:2014xxa} for its superspace construction.}
This multiplet is obtained by coupling the standard Weyl multiplet to on-shell vector and linear multiplets. The vector multiplet consists of a scalar field $\sigma$, the gaugino $\psi_{\alpha}^i$, an abelian gauge vector $C_{\mu}$ with field strength $G_{\mu \nu}= 2 \partial_{[\mu} C_{\nu]}$, and an $\rm{SU(2)}$ triplet of auxiliary fields $Y^{ij} = Y^{(ij)}$. The linear multiplet contains an $\rm{SU(2)}$ triplet of scalars $L^{ij} = L^{(ij)}$, a gauge three-form $E_{\mu \nu \rho}$, a scalar $N$, and an $\rm{SU(2)}$ doublet $\varphi^i_{\alpha}$. In the gauged dilaton Weyl multiplet, the independent gauge fields remain the same as the standard Weyl, but the matter content is replaced with $\{\sigma, C_\mu, B_{\mu\nu}, L_{ij}, E_{\mu \nu \rho}, N, \psi^i, \varphi^i\}$. The bosonic matter fields of the vector and the standard Weyl multiplet are then expressed as follows \cite{Coomans:2012cf}
	\begin{eqnarray}
	Y^{ij} &=&   - \frac{g}{2} \sigma^{-1} L^{ij} + {\rm f.t.}~, 
 \nonumber\\
	T_{ab} &=& \frac{1}{8}\sigma^{-1} G_{ab} + \frac{1}{48} \sigma^{-2} \epsilon_{abcde} H^{cde} + {\rm f.t.}~, 
 \nonumber\\
	D &=&  \frac{1}{4} \sigma^{-1} \nabla^a \nabla_a \sigma + \frac{1}{8} \sigma^{-2} (\nabla^a \sigma) \nabla_a \sigma -\frac{1}{32} R
 \nonumber\\
	&& - \frac{1}{16} \sigma^{-2} G^{ab} G_{ab} - ( \frac{26}{3}  T^{ab} -2 \sigma^{-1} G^{ab} )  T_{ab}
	\nonumber\\
	&& + \frac{g}{4} \sigma^{-2} N   + \frac{g^2}{16} \sigma^{-4} L^2 + {\rm f.t.}~,
 \label{dilaton weyl basis}
	\end{eqnarray}
	where ``f.t." stands for omitted fermionic terms and $H_{abc} = e_{a}{}^{\mu} e_{b}{}^{\nu} e_{c}{}^{\rho} H_{\mu \nu \rho}$ denotes the three-form field strength $H_{\mu \nu \rho}:= 3 \partial_{[\mu}B_{\nu \rho]}+\frac{3}{2}C_{[\mu}G_{\nu \rho]} + \frac{1}{2} g  E_{\mu \nu \rho}$.	In the above, the covariant derivative is denoted by
	\begin{eqnarray}
	\nabla_a = e_{a}{}^{\mu} \big( \partial_{\mu}-  \omega_{\mu}{}^{b c} M_{bc}- b_{\mu} \mathbb{D} - V_{\mu}{}^{i j} U_{i j} \big)~,
	\end{eqnarray}
	with $M_{ab}$, $\mathbb{D}$, and $U_{ij}$ being the Lorentz, dilatation, and SU(2) generators, respectively. The dilatation connection $b_{\mu}$ is pure gauge and will be set to zero throughout. The mapping \eqref{dilaton weyl basis} allows us to easily convert every invariant involving a coupling to the standard Weyl multiplet to that written in terms of the gauged dilaton Weyl multiplet. The ungauged map and the models can simply be obtained by setting $g=0$ in \eqref{dilaton weyl basis}. In this case, the fields of the linear multiplet decouple from the map \eqref{dilaton weyl basis}, and the multiplet reduces to the ungauged dilaton Weyl multiplet with $32+32$ off-shell degrees of freedom \cite{Bergshoeff:2001hc, Fujita:2001kv}.

\subsection{Curvature-squared actions in the gauged dilaton Weyl background}

\subsubsection{BF action}

 In the superconformal tensor calculus, the so-called BF action principle plays a fundamental role in the construction of general supergravity-matter couplings, see \cite{Butter:2014xxa,Kugo:2000af,Fujita:2001kv,Kugo:2002vc,Bergshoeff:2001hc,Bergshoeff:2002qk,Bergshoeff:2004kh} for the 5D case. It is based on an appropriate product of a linear multiplet with an Abelian vector multiplet:
	\begin{eqnarray}
	&&e^{-1} \mathcal{L}_{\rm BF} = \,A_{{a}} E^{{a}} + \rho N + \mathcal{Y}_{i j} L^{i j} + \rm f.t.~.
	\label{BF-Scomp}
	\end{eqnarray}
	Here we use $\{\rho, A_\mu, \mathcal{Y}_{ij}, \lambda^i_\alpha\}$ to denote the field content in an arbitrary vector multiplet, and the bosonic part of the constrained vector $E_a$ is related to the three-form gauge field $E_{abc}$ via $E_a = -\frac{1}{12} \epsilon_{abcde} \nabla^b E^{cde}$. Furthermore, in our construction we consider the case in which the linear multiplet fields $L^{ij}, N$ and $E_a$ are composite of the vector multiplet. The BF-action thus yields a vector-coupled action in the (gauged) dilaton Weyl background. Using the off-shell map given in \cite{Ozkan:2013uk}, a vector multiplet can be identified with fields in the gauged dilaton Weyl multiplet as
\begin{align}
	{\mathcal Y}^{ij} & \to \frac{1}{4} {\rm i}  \sigma^{-1} \bar\psi^i \psi^j - \frac{g}{2}  \sigma^{-1} L^{ij}\,, & \rho & \to \sigma\,, \nonumber\\
	 A_\mu & \to C_\mu \,, & \lambda^i & \to \psi^i \,,
	\label{VecToDW2}
\end{align}	
	which gives rise to off-shell models that are purely expressed in terms of the fields of the (gauged) dilaton Weyl multiplet. By appropriately choosing primary composite linear multiplets, eq. \eqref{BF-Scomp} becomes the building block for constructing various curvature-squared invariants.

 Within the superconformal approach, supersymmetric completions of the Weyl tensor squared and the Ricci scalar squared were constructed for the first time, respectively in \cite{HOT} and \cite{Ozkan:2013nwa}. 
The third independent invariant necessary to obtain all the curvature-squared models in five dimensions includes the Ricci tensor-squared term. In the standard Weyl multiplet background, this invariant was defined in superspace in \cite{Butter:2014xxa}.

\subsubsection{Weyl$^2$}

As an example, we first consider the composite linear multiplet superfield that is used to construct the off-shell Weyl tensor squared invariant. It is a composite of the standard Weyl multiplet superfields, that is
\begin{eqnarray}
L^{ij}_{\rm Weyl} \to H^{ij}_{\rm Weyl}:= - \frac{{{\rm i}}}{2}   W^{{\alpha} {\beta} {\gamma} i} W_{{\alpha} {\beta} {\gamma}}\,^{j}+\frac{3{{\rm i}}}{2}   W^{{\alpha} {\beta}} X_{{\alpha} {\beta}}\,^{i j} - \frac{3 {{\rm i}}}{4}   X^{{\alpha} i} X^{j}_{{\alpha}}~,
\label{primary-Weyl}
\end{eqnarray}
together with its (bosonic) descendants which are defined by 
\begin{subequations} \label{descendants-Weyl}
\begin{align}
 N_{{\rm Weyl}} &=  \frac{{{\rm i}}}{12} \nabla^{\alpha}_{i} \nabla_{\alpha j}H^{i j}_{\rm{Weyl}}|~, \\
E^{{a}}_{{\rm Weyl}} &= \frac{{{\rm i}}}{12} (\Gamma^{{a}})^{\alpha \beta} \nabla_{\alpha i} \nabla_{\beta j} H^{i j}_{\rm{Weyl}}|~,
\end{align}
\end{subequations}
with $\nabla_\alpha^{i}$ being the conformal superspace spinor derivative. In the above, the primary component fields $L^{ij}_{\rm Weyl}, N_{\rm Weyl}, E^{a}_{\rm Weyl}$ are obtained by bar-projection, which corresponds to setting $\theta=0$, see \cite{Butter:2014xxa,Gold:2023dfe} for details. 

In deriving the off-shell Weyl-squared invariant, one may start by first substituting the resulting composite multiplet \eqref{primary-Weyl} and \eqref{descendants-Weyl} into \eqref{BF-Scomp}. This yields the explicit form of the invariant written in the standard Weyl basis. Its corresponding form in the gauged dilaton Weyl background can be obtained by making use of the map \eqref{dilaton weyl basis} and \eqref{VecToDW2}. In this paper, we will only present the bosonic sector of the off-shell Weyl-squared invariant in the gauged dilaton Weyl background, and upon imposing the gauge conditions
	\begin{eqnarray}
	\sigma=1~, \qquad b_{\mu}=0~, \qquad \psi^{i} = 0~.
	\label{gauge}
	\end{eqnarray}
The action is given by
\begin{align}\label{Weyl-squared-gauged}
    e^{-1}\mathcal{L}_{{\textrm Weyl}^2} 
    =&  
    - \frac{1}{8} \epsilon^{{a} {b} {c} {d} {e}} C_{{a}} R_{{b} {c} f g} R_{{d} {e}}{}^{f g} 
    + \frac{1}{6} \epsilon^{{a} {b} {c} {d} {e}} C_{{a}} V_{{b} {c}}\,^{i j} V_{{d} {e} i j} 
    +\frac{2}{3} V^{{a} {b} i j} V_{{a} {b} i j} 
    \nonumber \\
    &  
    - \frac{1}{4}R_{{a} {b} {c} {d}} R^{{a} {b} {c} {d}} 
    +\frac{1}{3}R_{{a} {b}} R^{{a} {b}}  
    - \frac{1}{12}{R}^{2}
     \nonumber \\
    &
    + \frac{1}{3}R_{{a} {b} {c} {d}} ({{G}}^{{a} {b}} G^{{c} {d}} 
    - 2 H^{{a} {b}} H^{{c} {d}} -3 {{H}}^{{a} {b}} {{G}}^{{c} {d}}) 
    \nonumber \\
    &     
    - \frac{4}{3} R_{{a} {b}} {{H}}^{{a} {c}} G^{{b}}{}_{{c}}  
    +\frac{16}{3}R^{{a} {b}} {{H}}^{2}_{{a} {b}} 
    + \frac{1}{3}R {{H}}_{{a} {b}} G^{{a} {b}} 
     -\frac{4}{3}R H^2
    - 4(H^2)^2 
        \nonumber \\
    & 
     -8H^4
    -\frac{16}{3} H^2 H_{{c} {d}} G^{{c} {d}} 
    -\frac{40}{3}H^{2}_{{a} {d}} H^{a}{}_{ {c}}  G^{{c} {d}}
    +\frac{8}{3}H^2 G^2
    \nonumber \\
    &  
    +\frac{2}{3} H_{{a} {b}} H_{{c} {d}} G^{{a} {b}} G^{{c} {d}}    
    + \frac{1}{12}(G^2)^2 
    \nonumber \\
    &
    - \frac{16}{3}H^{2}_{{a} {b}}  G^{2 {a} {b}}
    - \frac{4}{3} H_{{a} {b}} H_{{c} {d}} G^{{a} {c}} G^{{b} {d}}
    - \frac{1}{3}H_{{a} {b}} G^{{a} {b}} G^{2}  
    +2 G^{2 {a} {b}} G^{{c}}{}_{{b}} H_{{c} {a}}    \nonumber \\
    &     
    -\frac{1}{3} (\nabla^{{a}}{G_{{b} {c}}}) \nabla_{{a}}{G^{{b} {c}}} 
     +\frac{8}{3} (\nabla^{{a}} H_{{b} {c}} )\nabla_{{a}} H^{{b} {c}} 
    - \frac{1}{2}G^{4}
    \nonumber \\
    &  
    +\frac{4}{3}\epsilon^{{a} {b} {c} {d} {e}} H_{{a} {b}} H_{{c} {d}} \nabla^{{f}}{G_{{e} {f}}}
    - 2 \epsilon^{{a} {b} {c} {d} {e}}  H_{{b} {f}} (\nabla_{{a}}H_{{c}}{}^{{f}}) G_{{d} {e}}
    \nonumber \\
    &  
    - \frac{2}{3}\epsilon^{{a} {b} {c} {d} {e}} H_{{a} {b}} (\nabla^{{f}}{G_{{c} {f}}} ) G_{{d} {e}}
    -\frac{1}{24}\epsilon^{{a} {b} {c} {d} {e}} (\nabla^{{f}}{G_{{a} {f}}}) G_{{b} {c}} G_{{d} {e}} 
    \nonumber \\
    & -g \Big( -\frac{4}{3} N H_{a b} G^{a b}  + \frac{4}{3} N G^{2} -8 N H^{2}
     +\frac{4}{3} V_{a b}{}^{ij} L_{ij} H^{a b} 
       \nonumber \\
    &~~~~~~~~~
- \frac{2}{3}V_{a b}{}^{ij} L_{ij} G^{a b} 
- \frac{2}{3} R N \Big)
    \nonumber\\
    &+ g^2 \left( \frac{1}{3}L^2 H^{a b} G_{a b} -\frac{1}{3}L^2 G^{2} +2 L^2 H^{2} -\frac{8}{3}N^2 + \frac{1}{6} R L^2\right) 
        \nonumber\\
    &
-\frac{4}{3} g^3 N L^2 
-\frac{1}{6} g^4 L^4
~,
\end{align}
where $V_{ab}{}^{ij} = 2 e_a{}^\mu e_b{}^\nu\partial_{[\mu} V_{\nu]}{}^{ij} - 2 V_{[a}{}^{k(i} V_{b]k}{}^{j)}$.  Furthermore, we have used the following notations: $H^{ab}=- \frac{1}{12}\epsilon^{abcde}H_{cde}$, $H^2 = H^{ab}H_{ab}$, $G^2 = G^{ab}G_{ab}$, $H^2_{ab}:=H_{a}{}^cH_{b c}$, $G^2_{ab}:=G_{a}{}^c G_{b c}$, $G^4 = G^{2 ab}G^{2}_{a b}$, and $H^4 = H^{2 ab}H^{2}_{a b}$.

\subsubsection{Log}
In superspace, the composite linear superfield associated with the ``Log multiplet" is described by
	\begin{equation}
		L^{ij}_{\rm Log} \rightarrow H^{i j}_{{\rm Log}} 
  :=  \frac{3 {{\rm i}}}{1280} \nabla^{(i j} \nabla^{k l)} \nabla_{k l} \log{{W}}~.
		\label{logW-0}
	\end{equation}
 Here, the primary superfield $H^{i j}_{{\rm Log}}$ is obtained by making use of the standard Weyl multiplet and by acting with six spinor covariant derivatives on $W$, with ${W}$ being the superfield describing the primary field strength of a compensating vector multiplet. In \eqref{logW-0}, we have denoted $\nabla^{ij} = \nabla^{\alpha (i} \nabla_{\alpha}^{j)}$, and the primary field $L^{i j}_{{\rm Log}} =H^{i j}_{{\rm Log}}|_{\theta=0} $ is obtained by projection to the lowest $\theta=0$ component.
 The rest of the composite Log multiplet is then obtained by further acting with up to two more spinor covariant derivatives on $H^{i j}_{{\rm Log}}$.	Due to the complexity of computing up to eight supersymmetry transformations, the explicit form of the  Log multiplet, including all fermionic terms, has been obtained only recently with the aid of the \textit{Cadabra} software \cite{Cadabra-1,Cadabra-2}. 
The expression of the lowest component, \eqref{logW-0}, in its expanded form in terms of the descendants of $W$ and $W_{\alpha \beta}$ was obtained in \cite{Gold:2023dfe}. It takes the form
\begin{eqnarray}
\label{HlogW-full}
    H^{i j}_{{\rm Log}}
    &=&
    {}  \frac{1}{2}  W_{{a} {b}}   \Phi^{a b i j}  - \frac{272{{\rm i}}}{3} \chi^{{\alpha} i} \chi^{j}_{{\alpha}} 
    \nonumber\\
    &&
    +{{W}}^{-1} \Bigg\{\,-6X^{i j} D + \frac{1}{2}  F_{{a} {b}} \Phi^{{a}{b} i j}  - \frac{1}{2}  \Box{{X^{i j}}} - \frac{9}{64}X^{i j} W^{{a} {b}} W_{{a} {b}} \nonumber\\
    &&~~~~~~~~~~
+ \frac{1}{4} W^{{a}{b}} W_{{a}{b}}{}^{\alpha (i} \lambda^{j)}_{\alpha} 
+ 8 {\rm i} (\Gamma^{{a}})^{{\alpha} {\beta}}   \chi^{(i}_{{\alpha}}  {\nabla}_{{a}}{\lambda^{j)}_{{\beta}}} - 8 {\rm i} (\Gamma^{{a}})^{{\alpha} {\beta}}   \lambda^{(i}_{{\alpha}} {\nabla}_{{a}}{\chi^{j)}_{{\beta}}}\Bigg\}
    \nonumber\\
    &&
    + {{W}}^{-2} \Bigg\{\,\frac{1}{2} X^{i j}  \Box{{{W}}}+\frac{1}{2}  \big({\nabla}^{{a}}{{W}} \big) {\nabla}_{{a}}{X^{i j}} 
+ \frac{1}{4} F^{{a}{b}} W_{{a}{b}}{}^{\alpha (i} \lambda^{j)}_{\alpha}-\frac{{\rm i}}{2}  \lambda^{{\alpha}(j}  \Box{{\lambda^{i)}_{{\alpha}}}} \nonumber\\
&&~~~~~~~~~~~
-\frac{{\rm i}}{4}    \big({\nabla}^{{a}}{\lambda^{{\alpha} i}} \big) {\nabla}_{{a}}{\lambda^{j}_{{\alpha}}} - \frac{3 {\rm i}}{16} \frak{\epsilon}^{{a} {b} {c} {d} {e}} (\Sigma_{{a} {b}})^{{\alpha} {\beta}}   W_{{d} {e}} \lambda^{(i}_{{\alpha}}  {\nabla}_{{c}}{\lambda^{j)}_{{\beta}}}
- 2 {{\rm i}}  D \lambda^{{\alpha} i} \lambda^{j}_{{\alpha}}
\nonumber\\
&&~~~~~~~~~~~
- \frac{3 {\rm i}}{8} (\Gamma^{{a}})^{{\alpha} {\beta}}  \lambda^{i}_{{\alpha}} \lambda^{j}_{{\beta}}  {\nabla}^{{c}}{W_{{a} {c}}} 
- 4 {{\rm i}} (\Sigma_{{a} {b}})^{{\alpha} {\beta}}  F^{{a} {b}} \chi^{(i}_{{\alpha}} \lambda^{j)}_{{\beta}}  
+ 4 {{\rm i}} X^{i j} \chi^{{\alpha} k} \lambda_{{\alpha} k}
\nonumber\\
&&~~~~~~~~~~~
+\frac{9 {{\rm i}}}{256} \frak{\epsilon}^{{a} {b} {c} {d} {e}} (\Gamma_{a})^{{\alpha} {\beta}}   W_{{b} {c}} W_{{d} {e}} \lambda^{i}_{{\alpha}} \lambda^{j}_{{\beta}} 
+ \frac{3 {{\rm i}}}{128} W^{{a} {b}} W_{{a} {b}} \lambda^{{\alpha} i} \lambda^{j}_{{\alpha}} 
\Bigg\} 
\nonumber\\
&&
+ {{W}}^{-3} \Bigg\{\,\frac{1}{8}X^{i j} F^{{a} {b}} F_{{a} {b}}  - \frac{1}{8} X^{i j} X^{k l} X_{k l} - \frac{1}{4} X^{i j}   \big({\nabla}^{{a}}{{W}} \big) {\nabla}_{{a}}{{W}}  \nonumber\\
&&~~~~~~~~~~
- \frac{{{\rm i}}}{8} \frak{\epsilon}^{{a} {b} {c} {d} {e}} (\Sigma_{{a} {b}})^{{\alpha} {\beta}}   F_{{d} {e}} \lambda^{(i}_{{\alpha}}  {\nabla}_{c}{\lambda^{j)}_{{\beta}}} - \frac{{{\rm i}}}{4} (\Gamma^{{a}})^{{\alpha} {\beta}}  F_{{a} {b}} \lambda^{(i}_{{\alpha}}  {\nabla}^{{b}}{\lambda^{j)}_{{\beta}}}  
\nonumber\\
&&~~~~~~~~~~
- \frac{{{\rm i}}}{4} (\Gamma^{{a}})^{{\alpha} {\beta}}   \lambda^{i}_{{\alpha}} \lambda^{j}_{{\beta}}  {\nabla}^{{c}}{F_{{a} {c}}} 
+\frac{{{\rm i}}}{4} (\Gamma^{{a}})^{{\alpha} {\beta}}  X^{i j} \lambda^{k}_{{\alpha}} {\nabla}_{{a}}{\lambda_{{\beta} k}}
\nonumber\\
&&~~~~~~~~~~
+\frac{{{\rm i}}}{4} (\Gamma^{{a}})^{{\alpha} {\beta}}  X^{k (i} \lambda^{j)}_{{\alpha}}  {\nabla}_{{a}}{\lambda_{{\beta} k}} 
- \frac{{{\rm i}}}{4} (\Gamma^{{a}})^{{\alpha} {\beta}}  \lambda^{(i}_{{\alpha}}    \big({\nabla}_{{a}}{X^{j) l}} \big) \lambda_{{\beta} l}
 \nonumber\\
&&~~~~~~~~~~
+ \frac{{{\rm i}}}{4}   \lambda^{{\alpha} i} \lambda^{j}_{{\alpha}}  \Box{{{W}}} 
+ \frac{3 {{\rm i}}}{4}   \big({\nabla}^{{a}}{{W}} \big) \lambda^{{\alpha} (i}   {\nabla}_{{a}}{\lambda^{j)}_{{\alpha}}} 
- \frac{{{\rm i}}}{2} (\Sigma_{{a} {b}})^{{\alpha} {\beta}}  \big({\nabla}^{{a}}{{W}}  \big)\lambda^{(i}_{{\alpha}}   {\nabla}^{{b}}{\lambda^{j)}_{{\beta}}}  
 \nonumber\\
 &&~~~~~~~~~~
  - \frac{3 {{\rm i}}}{16} (\Gamma^{{a}})^{{\alpha} {\beta}}  W_{{a} {b}} \lambda^{i}_{{\alpha}} \lambda^{j}_{{\beta}}  {\nabla}^{{b}}{{W}} 
  + \frac{3 {{\rm i}}}{32} W_{{a} {b}} F^{{a} {b}} \lambda^{{\alpha} i} \lambda^{j}_{{\alpha}} 
 \nonumber\\
 &&~~~~~~~~~~
 +\frac{9 {{\rm i}}}{64} \frak{\epsilon}^{{a} {b} {c} {d} {e}} (\Gamma_{{a}})^{{\alpha} {\beta}}   W_{{b} {c}} F_{{d} {e}} \lambda^{i}_{{\alpha}} \lambda^{j}_{{\beta}}  - \frac{3 {{\rm i}}}{32}  (\Sigma_{{a} {b}})^{{\alpha} {\beta}}  X^{i j} W^{{a} {b}} \lambda^{k}_{{\alpha}} \lambda_{{\beta}k}  \Bigg \}
 \nonumber\\
 &&
 + {{W}}^{-4}\Bigg\{  -\frac{3 {{\rm i}}}{16}   \lambda^{{\alpha} i} \lambda^{j}_{{\alpha}}   \big({\nabla}^{{a}}{{W}} \big) {\nabla}_{{a}}{{W}} 
 - \frac{3 {{\rm i}}}{8} (\Gamma^{{a}})^{{\alpha} {\beta}}   X^{ k (i} \lambda^{j)}_{{\alpha}} \lambda_{{\beta}k}  {\nabla}_{{a}}{{W}} \nonumber\\
 &&~~~~~~~~~~~
 +\frac{3{{\rm i}}}{8} (\Gamma^{{a}})^{{\alpha} {\beta}}  F_{{a} {b}} \lambda^{i}_{{\alpha}} \lambda^{j}_{{\beta}}  {\nabla}^{{b}}{{W}} 
 + \frac{3 {{\rm i}}}{32} F^{{a} {b}} F_{{a} {b}} \lambda^{{\alpha} i} \lambda^{j}_{{\alpha}} 
 \nonumber\\
 &&~~~~~~~~~~~
 +\frac{3{{\rm i}} }{64} \frak{\epsilon}^{{a} {b} {c} {d} {e}} (\Gamma_{{a}})^{{\alpha} {\beta}}   F_{{b} {c}} F_{{d} {e}} \lambda^{i}_{{\alpha}} \lambda^{j}_{{\beta}} 
 - \frac{3 {{\rm i}}}{16} (\Sigma_{{a} {b}})^{{\alpha} {\beta}}  X^{i j} F^{{a} {b}} \lambda^{k}_{{\alpha}} \lambda_{{\beta} k}  
 \nonumber\\
 &&~~~~~~~~~~~
 -\frac{3 {{\rm i}}}{16}  X^{i j} X^{k l} \lambda^{\alpha}_{k} \lambda_{{\alpha} l}   -\frac{3 {{\rm i}}}{32}  X^{k l} X_{ k l} \lambda^{{\alpha} i} \lambda^{j}_{{\alpha}}  
 \Bigg\}
  \nonumber\\
  &&
  +~ {\rm higher~fermions}
~.
\end{eqnarray}

We have suppressed the higher-fermion terms, as they do not contribute to the bosonic sector. Inserting the resulting composite multiplet into \eqref{BF-Scomp} yields the explicit form of a new ``Log invariant'' which was presented in \cite{Gold:2023ykx} in the standard Weyl basis. Then, the Log invariant in the gauged dilaton Weyl background can be obtained by employing the map \eqref{dilaton weyl basis} and \eqref{VecToDW2}. In this paper, it suffices to present its bosonic sector in the gauge  \eqref{gauge}. The gauged Log  invariant in the dilaton Weyl background, which includes a Ricci-squared term, reads
\begin{align}
    e^{-1}\mathcal{L}_{{\textrm{Log}}} =& 
    - \frac{1}{6} R_{{a} {b}} R^{{a} {b}}
    +\frac{1}{24}R^2 
    +\frac{1}{6}R^{{a} {b}}  G^{2}_{{a} {b}} + \frac{1}{3}R H_{{a} {b}} G^{{a} {b}}
    -\frac{4}{3}R_{{a} {b}}  H^{{a} {c}} G^{{b}}{}_{{c}} 
    \nonumber\\&
    -\frac{1}{3}R H^2
    - \frac{1}{12} \epsilon^{{a} {b} {c} {d} {e}} C_{{a}} V_{{b} {c}}\,^{i j} V_{{d} {e} i j} 
    +\frac{1}{6} V^{{a} {b} i j} V_{{a} {b} i j}
    - 2 ({{H}}^2)^2 
    \nonumber \\ &
    +\frac{16}{3}H^{2}_{{a} {b}}  H^{{a}{c}} G^{b}{}_{c} - \frac{4}{3}H^2 H_{{a} {b}} G^{{a} {b}}
    +\frac{2}{3} H_{{a}{b}} H_{{c} {d}} \big( G^{{a} {b}} G^{{c} {d}} -2  G^{{a} {c}} G^{{b} {d}} \big)
    \nonumber\\ &
    +\frac{2}{3}H^2 G^2
    -\frac{4}{3} H^{{2}{a} {b}}G^{2}_{{a} {b}}  - \frac{1}{3}H_{{a} {b}} G^{{a}{b}} G^2
    +G^{2}_{{a} {b}} H^{a c} G^{b}{}_{c}  - \frac{1}{48}(G^2)^2 
    \nonumber \\ &
    -\frac{1}{24} G^4 
    -\frac{1}{6} \nabla_{{c}}G^{{a} {c}}  \nabla^{{b}}G_{{a} {b}}
    +2 \nabla_{{a}}H_{{b} {c}}   \nabla^{[{a}}H^{{b} {c}]}
    \nonumber\\ &
    + \frac{1}{48}\epsilon^{{a} {b} {c} {d} {e}} \nabla^{{f}}G_{{e} f}  (4 H_{{a} {b}} 
    - G_{ {a} {b}}) (4 H_{{c} {d}} - G_{ {c} {d}}) 
    \nonumber\\ & 
    + \frac{g}{6}  \Big(   R N  -4 N H_{{a} {b}} G^{{a} {b}} - 2 N G^2 +  V_{{a} {b}}\,^{i j} L_{ij} ( G^{{a} {b}} + 4 H^{{a} {b}} )
    +12 N H^2 
    \nonumber\\
    &~~~~~~~~~
    - 6 \nabla^{{a}} \nabla_{{a}}N \Big)
    \nonumber\\&
    -  \frac{g^2}{24} \Big( 2 R L^2 -  L^2  (G^{2} -4 G^{{a} {b}}H_{{a} {b}} - 24  H^2 )  +4 N^2 +6 \nabla^{{a}}L^{ij}\nabla_{{a}}L_{ij}  \Big)
    \nonumber \\ &
    +\frac{2}{3}N {L}^{2} {g}^{3}+\frac{5}{24}{L}^{4} {g}^{4}~. \label{log-gauged}
\end{align}

\subsubsection{Scalar curvature squared}
In order to construct the supersymmetric completion of the Ricci scalar-squared invariant, one may start by defining a composite vector multiplet in terms of a linear multiplet. This composite vector multiplet is then substituted into the vector multiplet action obtained using \eqref{BF-Scomp}. The bosonic sector of the gauged scalar curvature-squared invariant in the gauged dilaton Weyl background, \eqref{dilaton weyl basis} and in the gauge \eqref{gauge}, is given by:
\begin{subequations}
    \begin{align}
        e^{-1}  \mathcal{L}_{R^2} =&  \mathbf{Y}^{i j} \mathbf{Y}_{i j} -  2\nabla^{{a}}(N L^{-1})\nabla_{{a}}(N L^{-1}) \nonumber \\ & -\frac{1}{8}  \epsilon_{{a} {b} {c} {d} {e}}C^{{a}}\mathbf{G}^{{b} {c}}\mathbf{G}^{{d} {e}} + N L^{-1} G^{{a} {b}}  \mathbf{G}_{{a} {b}}  -   N^2 L^{-2} G^{a b} G_{a b}    \nonumber \\
    &+ 4N^2 L^{-2} H^{a b}G_{a b}     - \frac{1}{4} \mathbf{G}^{{a} {b}}\mathbf{G}_{{a} {b}}  -4 N L^{-1}  H_{a b}  \mathbf{G}^{{a} {b}}  \nonumber \\
   &
    +  g^2  \Big(    \frac{1}{4} L^{ij} \nabla^{a} \nabla_{a}{L_{ij}} 
    -  \frac{1}{4}R L^2
-  H^{2} L^2
 +  \frac{1}{8} G^{2} L^2 - \frac{5}{2}  N^2 
 -  \frac{1}{2} E^{a}E_{a}  
  \nonumber\\
    &~~~~~~~~~
    - \frac{1}{2} \nabla^{a}{L} \nabla_{a}{L}
\Big) 
    \nonumber\\
    &
    - 4 g   N^3 L^{-2}  + \frac{1}{16} g^4   L^4 ~,
     \label{R2-eq}
    \end{align}
    where,
    \begin{align}
        \mathbf{G}_{{a} {b}} 
    =&   4 \nabla_{[a} ( L^{-1} E_{b]}) + 2 L^{-1} L_{ij}(V_{ab}{}^{ij})
    -2 L^{-3}  L_{ij}  \big(\nabla_{[a} L^{ik} \big) \nabla_{b]} L_k{}^{j}~,
    \nonumber\\ 
        \mathbf{Y}^{ij}{} 
    =&   \frac{1}{4} L^{-1} \left( 4 \nabla^{a} \nabla_{a}{L^{ij}} - 2 R {L^{ij}}
    - 8
     H^{2} {L^{ij}}
     +   G^{2} {L^{ij}}
    \right)
    \nonumber\\
    &+  L^{-3} \left( - N^2 {L^{ij}} -  E^{a}E_{a} {L^{ij}} - 2 E^{a}L^{k(i} \nabla_{a} L_k{}^{j)}- {L_{kl}} \nabla^{a}  {L^{k(i}} \nabla_{a} {L^{j)l}}
    \right) \ .
    \end{align}
\end{subequations}

\subsection{Gauged Riemann-squared Action}
In this section, we construct the gauged Riemann-squared action. It turns out that the supersymmetric extension of the gauged Riemann-squared invariant can be achieved by taking a linear combination of the new curvature-squared ${\rm Log}$ invariant and the Weyl-squared invariant. By closely examining \eqref{log-gauged} and \eqref{Weyl-squared-gauged}, it becomes evident that the following combination
\begin{align}
    \mathcal{L}_{\rm Riem^2} = \mathcal{L}_{\rm Weyl^2} + 2 \mathcal{L}_{{\rm Log}}
\end{align}
provides an off-shell supersymmetric description of the gauged Riemann-squared invariant. Accordingly, we can define the corresponding composite linear multiplet $H^{ij}_{\rm Riem^2}$ for the Riemann-squared invariant as
\begin{align}
    H^{ij}_{\rm Riem^2} = H^{ij}_{\rm Weyl^2} + 2 H^{ij}_{{\rm Log}}~.
\end{align}
The resulting bosonic sector of gauged Riemann-squared Lagrangian is given by
\begin{align}
    e^{-1} \mathcal{L}_{{\rm Riem}^2} =&  
    - \frac{1}{8} \epsilon^{{a} {b} {c} {d} {e}} C_{{a}} R_{{b} {c} f g} R_{{d} {e}}{}^{f g}  
    + V^{{a} {b} i j} V_{{a} {b} i j} 
    \nonumber \\
    &  
    - \frac{1}{4}R_{{a} {b} {c} {d}} R^{{a} {b} {c} {d}} 
    + \frac{1}{3}R_{{a} {b} {c} {d}} ({{G}}^{{a} {b}} G^{{c} {d}} 
    - 2 H^{{a} {b}} H^{{c} {d}} -3 {{H}}^{{a} {b}} {{G}}^{{c} {d}}) 
    \nonumber \\
    &     
    - 4 R_{{a} {b}} {{H}}^{{a} {c}} G^{{b}}{}_{{c}}  
    +\frac{16}{3}R^{{a} {b}} {{H}}^{2}_{{a} {b}} 
    +\frac{1}{3}R^{{a} {b}}  G^{2}_{{a} {b}}
    + R {{H}}_{{a} {b}} G^{{a} {b}} 
     - 2 R H^2
        \nonumber \\
    & 
    - 8 (H^2)^2 
    -8H^4
    - 8 H^2 H_{{c} {d}} G^{{c} {d}} 
    - 24 H^{2}_{{a} {d}} H^{a}{}_{ {c}}  G^{{c} {d}}
    + 4 H^2 G^2
    \nonumber \\
    &  
    + 2 H_{{a} {b}} H_{{c} {d}} G^{{a} {b}} G^{{c} {d}}    
    + \frac{1}{24}(G^2)^2  - \frac{7}{12}G^{4}
    \nonumber \\
    &
    - 8 H^{2}_{{a} {b}}  G^{2 {a} {b}}
    - 4 H_{{a} {b}} H_{{c} {d}} G^{{a} {c}} G^{{b} {d}}
    - H_{{a} {b}} G^{{a} {b}} G^{2}  
    + 4 G^{2 {a} {b}} G^{{c}}{}_{{b}} H_{{c} {a}}    \nonumber \\
    &     
    -\frac{1}{3} (\nabla^{{a}}{G_{{b} {c}}}) \nabla_{{a}}{G^{{b} {c}}} 
     +\frac{8}{3} (\nabla^{{a}} H_{{b} {c}} )\nabla_{{a}} H^{{b} {c}} 
     \nonumber\\
    &
    -\frac{1}{3} \nabla_{{c}}G^{{a} {c}}  \nabla^{{b}}G_{{a} {b}}
    +4 \nabla_{{a}}H_{{b} {c}}   \nabla^{[{a}}H^{{b} {c}]} 
    \nonumber \\
    &  
    +2 \epsilon^{{a} {b} {c} {d} {e}} H_{{a} {b}} H_{{c} {d}} \nabla^{{f}}{G_{{e} {f}}}
    - 2 \epsilon^{{a} {b} {c} {d} {e}}  H_{{b} {f}} (\nabla_{{a}}H_{{c}}{}^{{f}}) G_{{d} {e}}
    \nonumber \\
    &  
    - \epsilon^{{a} {b} {c} {d} {e}} H_{{a} {b}} (\nabla^{{f}}{G_{{c} {f}}} ) G_{{d} {e}}
    \nonumber \\
   & - 2  g N G^2 +12  g N   {{H}}^2 +  g R N +  g V_{{a} {b}}\,^{i j}  L_{i j} G^{{a} {b}} 
    \nonumber
    \\ 
    &
    - 2 g \nabla^{{a}}{\nabla_{{a}}{N}} 
    - \frac{1}{4}{g}^{2}  G^2 {L}^{2} - \frac{1}{2}  g^{2} \nabla^{{a}}L^{i j} \nabla_{{a}} L_{i j}
     \nonumber
    \\ 
    &
    -  3   g^{2}{N}^{2}
    +\frac{1}{4}  g^{4} L^{4}
+{\rm fermions}~.
\end{align}
The above action (up to total derivative terms) can be compactly written as 
\begin{align}
    e^{-1} \mathcal{L}_{{\rm Riem}^2} =& - \frac{1}{4} \left( R_{\mu \nu a b} (\omega_{+}) - G_{\mu \nu} G_{a b}\right) \left(R^{\mu \nu a b}(\omega_{+}) - G^{\mu \nu}G^{a b} \right)
    \nonumber \\
    &
    - \frac{1}{2}\nabla_{\mu}(\omega_{+})G^{a b} \nabla^{\mu}(\omega_{+}) G_{a b} + V_{\mu \nu}{}^{i j} V^{\mu \nu}{}_{i j}
    \nonumber \\
    &
    - \frac{1}{8} \epsilon^{\mu \nu \rho \sigma \lambda} R_{\mu \nu a b}(\omega_{+})  R_{\rho \sigma}{}^{a b}(\omega_{+}) C_{\lambda}
    \nonumber \\
    &
    + \frac{1}{12} \epsilon^{\mu \nu \rho \sigma \lambda}H_{\rho \sigma \lambda}\left( 2 R_{\mu \nu a b} (\omega_{+}) - G_{\mu \nu} G_{a b}\right)  G^{a b}  
    \nonumber \\
    & - 2  g N G^2 +12  g N   {{H}}^2 +  g R N +  g V_{{a} {b}}\,^{i j}  L_{i j} G^{{a} {b}} 
    \nonumber
    \\ 
    &
    - 2 g \nabla^{{a}}{\nabla_{{a}}{N}} 
    - \frac{1}{4}{g}^{2}  G^2 {L}^{2} - \frac{1}{2}  g^{2} \nabla^{{a}}L^{i j} \nabla_{{a}} L_{i j}
     \nonumber
    \\ 
    &
    -  3   g^{2}{N}^{2}
    +\frac{1}{4}  g^{4} L^{4}+ {\rm fermions}~,
\end{align}
or
\begin{align}
    e^{-1} \mathcal{L}_{{\rm Riem}^2} =& - \frac{1}{4} \left( R_{\mu \nu a b} (\omega_{+}) - G_{\mu \nu} G_{a b}\right) \left(R^{\mu \nu a b}(\omega_{+}) - G^{\mu \nu}G^{a b} \right)
    \nonumber \\
    &
    - \frac{1}{2}\nabla_{\mu}(\omega_{+})G^{a b} \nabla^{\mu}(\omega_{+}) G_{a b} + V_{\mu \nu}{}^{i j} V^{\mu \nu}{}_{i j}
    \nonumber \\
    &
    - \frac{1}{8} \epsilon^{\mu \nu \rho \sigma \lambda} R_{\mu \nu a b}(\omega_{+})  R_{\rho \sigma}{}^{a b}(\omega_{+}) C_{\lambda}
    \nonumber \\
    &
    + \frac{1}{12} \epsilon^{\mu \nu \rho \sigma \lambda}H_{\rho \sigma \lambda}\left( 2 R_{\mu \nu a b} (\omega_{+}) - G_{\mu \nu} G_{a b}\right)  G^{a b}  
    \nonumber \\
    &- 2  g N G^2 +12  g N   {{H}}^2 +  g R N +  g V_{{a} {b}}\,^{i j}  L_{i j} G^{{a} {b}}  
    \nonumber
    \\ &
    - 2 g \nabla^{\mu}(\omega_{+})\nabla_{\mu}(\omega_{+}){N}
    - \frac{1}{4}{g}^{2}  G^2 {L}^{2} 
     \nonumber
    \\ &
    - \frac{1}{2}  g^{2} \nabla^{\mu}(\omega_{+})L^{i j} \nabla_{\mu}(\omega_{+}) L_{i j}
    -  3   g^{2}{N}^{2}
    +\frac{1}{4}  g^{4} L^{4}+ {\rm fermions}~,
\end{align}
where we have introduced the super-covariant curvature $ R_{\mu \nu a b} (\omega_{+})$, and derivative $\nabla^{\mu}(\omega_{+})$ associated with the torsionful spin connection $\omega_{+}$ as
\begin{subequations}
\begin{align}
    \omega_{\mu}{}^{ab}_{+} &=  \omega_{\mu}{}^{ab} + H_{\mu}{}^{ab}~, \\
    \nabla_{\mu}(\omega_{+}) &=  \nabla_{\mu} - \frac{1}{2} H_{\mu}{}^{ab} M_{ab}~, \\
    R_{\mu \nu a b} (\omega_{+}) &=   R_{\mu \nu a b} + 2 \nabla_{[\mu}(\omega_{+}) H_{\nu]}{}^{cd} 
    - 2 H_{[\mu}{}^{ce}H_{\nu]e}{}^{d}~.
\end{align}
\end{subequations}
In the ungauged limit, it is easy to see that our action aligns with those derived from the circle reduction of six-dimensional actions \cite{Bergshoeff:2011xn}. This can be seen by recognizing that in the ungauged limit the term involving the three-form can be equivalently expressed as:
\begin{align}
    &\frac{1}{12} \epsilon^{\mu \nu \rho \sigma \lambda}H_{\rho \sigma \lambda}\left( 2 R_{\mu \nu a b} (\omega_{+}) - G_{\mu \nu} G_{a b}\right)  G^{a b} \nonumber \\
   = & - \frac{1}{2} \epsilon^{\mu \nu \rho \sigma \lambda} B_{\rho \sigma}\left(R_{\mu \nu a b}(\omega_{+}) - G_{\mu \nu}G_{a b}\right) \nabla_{\lambda}(\omega_{+}) G^{a b} 
   \nonumber \\
   & + \frac{1}{8} \epsilon^{\mu \nu \rho \sigma \lambda} C_{ \lambda } G_{\rho \sigma} \left( 2 R_{\mu \nu a b} (\omega_{+}) - G_{\mu \nu} G_{a b}\right)  G^{a b} \nonumber \\
   & + \frac{1}{2} \nabla_{\lambda}(\omega_{+}) \left(\epsilon^{\mu \nu \rho \sigma \lambda} B_{\rho \sigma}\left(R_{\mu \nu a b}(\omega_{+}) - G_{\mu \nu}G_{a b}\right)  G^{a b}\right) ~,
\end{align}
where we have used the following identities:
\begin{subequations}
\begin{align}
    &\nabla_{[\lambda}(\omega_{+}) B_{\rho \sigma]} = \partial_{[\lambda} B_{\rho \sigma]} = \frac{1}{3} H_{\lambda \rho \sigma} +\frac{1}{2} C_{[\lambda} G_{\rho \sigma]}  ~, \\
    &\nabla_{[\lambda}(\omega_{+}) G_{\rho \sigma]} = \nabla_{[\lambda} G_{\rho \sigma]} = \partial_{[\lambda} G_{\rho \sigma]} = 0 ~, \\
    & \nabla_{[\lambda}(\omega_{+}) R_{\rho \sigma] a b} (\omega_{+}) = 0  ~.
\end{align}
\end{subequations}
Our gauged Riemann squared action is then a new result of this paper.

\subsection{Gauged Gauss-Bonnet Action}
In this section, we construct the supersymmetric extension of the gauged Gauss-Bonnet (GB) action. The GB term is a specific combination of curvature-squared terms,
\begin{align}
    - \frac{1}{4}R_{{a} {b} {c} {d}} R^{{a} {b} {c} {d}} 
    + R_{{a} {b}} R^{{a} {b}}  
    - \frac{1}{4}{R}^{2},
\end{align}
which is topological in four dimensions, meaning it does not contribute dynamically to the field equations. The GB combination, known for being ghost-free, plays a fundamental role in higher-derivative corrections and anomaly cancellation \cite{Zumino:1985dp, Metsaev:1987zx}. In fact, it is expected to control the leading-order $\alpha'$ corrections in string theory \cite{Zwiebach:1985uq, Deser:1986xr}. In five dimensions, the ungauged GB invariant has been constructed off-shell and fully classified \cite{Ozkan:2013uk, Ozkan:2013nwa, Butter:2014xxa}. However, the supersymmetric gauged GB invariant has remained absent from the literature. Interestingly, we find that the supersymmetric extension of the gauged GB invariant can be formulated as a linear combination of the new curvature-squared ${\rm Log}$ invariant and the Weyl-squared invariant. By closely examining \eqref{log-gauged} and \eqref{Weyl-squared-gauged}, it becomes evident that the following combination
\begin{align}
    \mathcal{L}_{\rm GB} = \mathcal{L}_{\rm Weyl^2} -4 \mathcal{L}_{{\rm Log}}~,
\end{align}
provides an off-shell supersymmetric description of the gauged GB invariant. Accordingly, we can define the corresponding composite linear multiplet $H^{ij}_{\rm GB}$ for the GB invariant as
\begin{align}
    H^{ij}_{\rm GB} = H^{ij}_{\rm Weyl^2} - 4 H^{ij}_{{\rm Log}}~.
\end{align}
The resulting gauged GB Lagrangian is given by
    \begin{align}
   e^{-1} \mathcal{L}_{\rm GB} =  &
    - \frac{1}{4}R_{{a} {b} {c} {d}} R^{{a} {b} {c} {d}} 
    + R_{{a} {b}} R^{{a} {b}}  
    - \frac{1}{4}{R}^{2} - \frac{1}{8} \epsilon^{{a} {b} {c} {d} {e}} C_{{a}} R_{{b} {c} f g} R_{{d} {e}}{}^{f g} 
     \nonumber \\
    &
    + \frac{1}{2} \epsilon^{{a} {b} {c} {d} {e}} C_{{a}} V_{{b} {c}}\,^{i j} V_{{d} {e} i j} 
    + \frac{1}{3}R_{{a} {b} {c} {d}} ({{G}}^{{a} {b}} G^{{c} {d}} 
    - 2 H^{{a} {b}} H^{{c} {d}} -3 {{H}}^{{a} {b}} {{G}}^{{c} {d}}) 
    \nonumber \\
    &     
    + 4 R_{{a} {b}} {{H}}^{{a} {c}} G^{{b}}{}_{{c}}  
    +\frac{16}{3}R^{{a} {b}} {{H}}^{2}_{{a} {b}} 
     -\frac{2}{3}R^{{a} {b}}  G^{2}_{{a} {b}} 
    - R {{H}}_{{a} {b}} G^{{a} {b}} 
    + 4(H^2)^2 
        \nonumber \\
    & 
     -8H^4
    + 8 H^{2}_{{a} {d}} H^{a}{}_{ {c}}  G^{{c} {d}}
    - 2 H_{{a} {b}} H_{{c} {d}} G^{{a} {b}} G^{{c} {d}}    
    + \frac{1}{6}(G^2)^2 
    \nonumber \\
    &
    + 4 H_{{a} {b}} H_{{c} {d}} G^{{a} {c}} G^{{b} {d}}
    + H_{{a} {b}} G^{{a} {b}} G^{2}  
    - 2 G^{2 {a} {b}} G^{{c}}{}_{{b}} H_{{c} {a}}    \nonumber \\
    &     
    -\frac{1}{3} (\nabla^{{a}}{G_{{b} {c}}}) \nabla_{{a}}{G^{{b} {c}}} 
     +\frac{8}{3} (\nabla^{{a}} H_{{b} {c}} )\nabla_{{a}} H^{{b} {c}} 
    - \frac{1}{3}G^{4}
    \nonumber \\ &
    +\frac{2}{3} \nabla_{{c}}G^{{a} {c}}  \nabla^{{b}}G_{{a} {b}}
    -8 \nabla_{{a}}H_{{b} {c}}   \nabla^{[{a}}H^{{b} {c}]}
    \nonumber \\
    &  
    - 2 \epsilon^{{a} {b} {c} {d} {e}}  H_{{b} {f}} (\nabla_{{a}}H_{{c}}{}^{{f}}) G_{{d} {e}} 
    -\frac{1}{8}\epsilon^{{a} {b} {c} {d} {e}} (\nabla^{{f}}{G_{{a} {f}}}) G_{{b} {c}} G_{{d} {e}} 
    \nonumber \\
    & + g \left( 4 N H_{a b} G^{a b}  
     - 4 V_{a b}{}^{ij} L_{ij} H^{a b}  + 4 \nabla^{{a}} \nabla_{{a}}N \right)
    \nonumber\\
    &+ g^2 \left( L^2 H^{a b} G_{a b} -\frac{1}{2}L^2 G^{2} +6 L^2 H^{2} - 2 N^2 + \frac{1}{2} R L^2  + \nabla^{{a}}L^{ij}\nabla_{{a}}L_{ij}\right) 
        \nonumber\\
    &
- 4 g^3 N L^2 
- g^4 L^4~.
\end{align}

\section{On-shell Analysis and holographic central charges}
\label{Holography-section-1}

In this section, we investigate the on-shell analysis of the quadratic curvature corrections to  $5D$  minimal gauged supergravity by applying proper field redefinitions, which allows us to  extract the central charges and obtain the corrected thermodynamic quantities. Based on previous work \cite{Reall:2019sah,Hu:2023gru,Ma:2024ynp},  we find that the computation of the leading higher curvature contributions to the Euclidean action of AdS black holes is drastically simplified once the higher curvature terms are expressed in terms of powers of the Weyl tensor. Due to the fact that the Weyl tensor vanishes sufficiently fast near the AdS boundary,  we can avoid some of the complications typically introduced by higher-derivative GHY terms or boundary counterterms. In the next section, we will illustrate how to compute the leading higher curvature contributions to thermodynamic quantities based on the redefined action.

\subsection{On-shell Analysis}\label{Section3.1}

The conformal supergravity in a gauged dilaton Weyl multiplet basis can be obtained by using the BF-action \eqref{BF-Scomp}, along with the composite definitions \eqref{dilaton weyl basis}. Imposing the gauge fixing conditions \eqref{gauge}, along with the following choice to break the $SU(2)_R$ symmetry to $U(1)_R$
\begin{equation}
  L_{ij}  = \frac{1}{\sqrt{2}} \delta_{ij} L\,,  
\end{equation}
leads to the gauged off-shell Poincar\'e supergravity
\begin{eqnarray}
	e^{-1} {\mathcal{L}} &=& L \left( R - \frac12 G_{ab} G^{ab} + 4 H_{ab} H^{ab} + 2 V_a^{\prime ij} V^{\prime a}_{ij} \right) \nonumber\\
	&& + L^{-1} \partial_a L \partial^a L - 2 L^{-1} E^a E_a - 2 \sqrt{2} E^a V_a - 2 L^{-1} N^2 \nonumber\\
	&& - 4 g E^a C_a - 2 g L N - 4 g N - \frac12 g^2 L^3 + 2 g^2 L^2 \,,
\end{eqnarray}
where we have split $V^{ij}_a$ into its traceless part $V^{'ij}_a$ and trace part $V_a$. Specically, 
\begin{equation}
 V_a^{ij} = V_a^{\prime ij} + \frac12 \delta^{ij} V_a\ .   
\end{equation}
From this Lagrangian, the auxiliary fields $N\,,  V_a^{\prime ij}$ can be eliminated by their equations of motion
\begin{equation}
	N  = - \frac12 g L (2+L) \,, \qquad V_a^{\prime ij}  = 0 \,. 
\end{equation}
We can, furthermore, dualize the two-form gauge field $B_{\mu\nu}$ to a vector field $\tilde C_\mu$ by adding a Lagrange multiplier term \cite{Coomans:2012cf}. The resulting model can be further truncated to a minimal model by setting 
\begin{equation}
    L  = 1 \,, \qquad \tilde C_a  = C_a \,.
\end{equation}
The field equation for $E_a$, then, leads to $V_a = -3g C_a/\sqrt{2} $. After eliminating all the auxiliary fields, we end up with the on-shell minimal gauged supergravity
\begin{eqnarray}
    e^{-1} {\mathcal{L}}_R &=& R+ 12g^2-\frac{1}{4}G^{\mu\nu}G_{\mu\nu}+\frac{1}{12\sqrt{3}} \epsilon^{\mu\nu\rho\sigma\delta}G_{\mu\nu}G_{\rho\sigma}C_{\delta}\ ,
\end{eqnarray}
where to get the standard kinetic term for the graviphoton, we rescaled $C_{\mu} \to C_{\mu}/\sqrt{3} $. We have also rescaled $g \rightarrow \sqrt{2} g$.

\subsubsection{Quadratic curvature corrections to 5D minimal gauged supergravity}

Owing to the off-shell nature of the construction, one can combine the 2- and 4- derivative supergravity actions without modifying the supertransformations. As discussed in \cite{Gold:2023ymc}, up to the first order in 4-derivative couplings, the 4-derivative supergravities can be put on-shell, by simply plugging in the gauge fixing conditions and solutions of the auxiliary fields implied by the 2-derivative action. The resulting on-shell action is parameterized as \cite{Gold:2023ymc,Ozkan:2024euj}
\begin{align}
	I_{5D,\:N=1} & =\frac{\sigma_{0}}{16\pi}\int\sqrt{-g}d^{5}x(\mathcal{L}_{0}+c_{1}\mathcal{L}_{{\rm Weyl}^{2}}+c_{2}\mathcal{L}_{{\rm Ricci}^2}+c_{3}\mathcal{L}_{R^{2}})\ ,\nonumber\\
	\mathcal{L}_{0} & =R+12\ell_{0}^{-2}-\frac{1}{4{\tilde g}_{0}^{2}}F^{\mu\nu}F_{\mu\nu}+\frac{1}{12\sqrt{3}{\tilde g}_{0}^{3}}\epsilon^{\mu\nu\rho\sigma\delta}F_{\mu\nu}F_{\rho\sigma}A_{\delta}\ , 
 \label{SUGRA}
\end{align}
where we introduced $\sigma_0=1/G$ for later convenience. We also redefined $g^2 \to 1/\ell_0^2$, and $C_{\mu}\to A_{\mu}/\tilde{g}_0$ where $\tilde{g}_0$ is a bookkeeping parameter which doesn't affect physical quantities. The 4-derivative action consists of three contributions $\mathcal{L}_{{\rm Weyl}^{2}}$,
 $\mathcal{L}_{{\rm Ricci}^2}$ and $\mathcal{L}_{R^{2}}$. The on-shell Weyl squared
action $\mathcal{L}_{{\rm Weyl}^{2}}$ is
\begin{align}
	\mathcal{L}_{{\rm Weyl}^{2}} & =R_{\mu\nu\rho\sigma}R^{\mu\nu\rho\sigma}-\frac{4}{3}R_{\mu\nu}R^{\mu\nu}+\frac{1}{3}R^{2}-\frac{1}{2\tilde{g}_{0}^{2}}R_{\mu\nu\rho\sigma}F^{\mu\nu}F^{\rho\sigma}\nonumber \\
	& 
	-\frac{4}{3\tilde{g}_{0}^{2}}R_{\mu\nu}F^{\mu\lambda}F_{\ \lambda}^{\nu}+\frac{2}{9\tilde{g}_{0}^{2}}RF^{\mu\nu}F_{\mu\nu}-\frac{2}{9\tilde{g}_{0}^{2}}\nabla^{\mu}F_{\mu\rho}\nabla_{\nu}F^{\nu\rho}-\frac{61}{432\tilde{g}_{0}^{4}}(F^{\mu\nu}F_{\mu\nu})^{2}\nonumber \\
	& 
	+\frac{5}{8\tilde{g}_{0}^{4}}F_{\mu\nu}F^{\nu\lambda}F_{\lambda\delta}F^{\delta\mu}+\frac{5\epsilon_{\mu\nu\rho\sigma\alpha}F^{\mu\nu}F^{\rho\sigma}\nabla_{\beta}F^{\beta\alpha}}{72\sqrt{3}\tilde{g}_{0}^{3}}+\frac{200}{3\ell_{0}^{4}}-\frac{35}{9\ell_{0}^{2}\tilde{g}_{0}^{2}}F^{\mu\nu}F_{\mu\nu}\nonumber \\
	& 
	+\frac{20}{3\ell_{0}^{2}}R-\frac{\epsilon^{\mu\nu\rho\sigma\alpha}F_{\mu\nu}F_{\rho\sigma}A_{\alpha}}{\sqrt{3}\ell_{0}^{2}\tilde{g}_{0}^{3}}+\frac{\epsilon^{\mu\nu\rho\sigma\alpha}A_{\mu}R_{\nu\rho}^{\ \ \ \beta\gamma}R_{\sigma\alpha\beta\gamma}}{2\sqrt{3}\tilde{g}_{0}}\ .
\end{align}
The on-shell Ricci tensor squared action $\mathcal{L}_{{\rm Ricci}^2}$ is
\begin{align}
	\mathcal{L}_{{\rm Ricci}^2} & =R^{2}-4R_{\mu\nu}R^{\mu\nu}-\frac{5}{6\tilde{g}_{0}^{2}}RF^{\mu\nu}F_{\mu\nu}-\frac{2}{\tilde{g}_{0}^{2}}\nabla^{\mu}F_{\mu\rho}\nabla_{\nu}F^{\nu\rho}+\frac{4}{\tilde{g}_{0}^{2}}R_{\mu\nu}F^{\mu\lambda}F_{\ \lambda}^{\nu}\nonumber \\
	& 
	+\frac{5}{16\tilde{g}_{0}^{4}}(F^{\mu\nu}F_{\mu\nu})^{2}-\frac{11}{9\tilde{g}_{0}^{4}}F_{\mu\nu}F^{\nu\lambda}F_{\lambda\delta}F^{\delta\mu}-\frac{2}{3\sqrt{3}\tilde{g}_{0}^{3}}\epsilon_{\mu\nu\rho\sigma\alpha}F^{\mu\nu}F^{\rho\sigma}\nabla_{\beta}F^{\beta\alpha}\nonumber \\
	& 
	-\frac{112}{\ell_{0}^{4}}+\frac{22}{3\ell_{0}^{2}\tilde{g}_{0}^{2}}F^{\mu\nu}F_{\mu\nu}-\frac{16}{\ell_{0}^{2}}R-\frac{\sqrt{3}}{\ell_{0}^{2}\tilde{g}_{0}^{3}}\epsilon^{\mu\nu\rho\sigma\alpha}A_{\mu}F_{\nu\rho}F_{\sigma\alpha}\ .
\end{align}
The on-shell Ricci scalar squared action $\mathcal{L}_{R^{2}}$ is
\begin{align}
\mathcal{L}_{R^{2}}&=R^{2}-\frac{1}{6\tilde{g}_{0}^{2}}RF^{\mu\nu}F_{\mu\nu}+\frac{1}{144\tilde{g}_{0}^{4}}(F^{\mu\nu}F_{\mu\nu})^{2}+\frac{16}{\ell_{0}^{2}}R \nonumber \\
	&+\frac{14}{3\ell_{0}^{2}\tilde{g}_{0}^{2}}F^{\mu\nu}F_{\mu\nu}-\frac{\sqrt{3}}{\ell_{0}^{2}\tilde{g}_{0}^{3}}\epsilon^{\mu\nu\rho\sigma\alpha}F_{\mu\nu}F_{\rho\sigma}A_{\alpha}+\frac{208}{\ell_{0}^{4}}\ .
\end{align}
\subsubsection{Field redefinitions and  redefined action}
The original on-shell actions of curvature squared supergravity in 5D minimal gauged supergravity are complicated. To simplify the discussion of black hole thermodynamics, we apply appropriate field redefinitions preserving black hole thermodynamics to express  the action in terms of powers of Weyl tensor. To proceed,  we first perform the field redefinitions
\begin{align}
	g_{\mu\nu} & \rightarrow g_{\mu\nu}'=g_{\mu\nu}+\lambda_{0}g_{\mu\nu}+\lambda_{1}R_{\mu\nu}+\lambda_{2}g_{\mu\nu}R+\frac{\lambda_{3}}{\tilde{g}_{0}^{2}}F_{\mu\nu}^{2}+\frac{\lambda_{4}}{\tilde{g}_{0}^{2}}g_{\mu\nu}F^{2}\ ,\nonumber \\
	A_{\nu} & \rightarrow A_{\nu}'=A_{\nu}+\lambda_{5}A_{\nu}+\lambda_{6}\nabla^{\rho}F_{\rho\nu}+\frac{\lambda_{7}}{\tilde{g}_{0}}\epsilon_{\nu\alpha\beta\rho\lambda}F^{\alpha\beta}F^{\rho\lambda}\ ,
 \label{RedefWeyl}
\end{align}
with the following parameters
\begin{align}
\lambda_{0}&=4(\lambda_{1}+5\lambda_{2})\ell_{0}^{-2},\ \ \lambda_{1}=-4c_{2},\ \ \lambda_{2}=\frac{1}{9}\big(-c_{1}+6(c_{2}-c_{3})\big),\ \ \lambda_{3}=2(c_{2}-c_{1})\ , \nonumber \\
\lambda_{4}&=\frac{1}{108}\left(49c_{1}+6(c_{3}-7c_{2})\right),\  \lambda_{5}=0,\  \lambda_{6}=\frac{2}{9}(c_{1}+9c_{2}),\  \lambda_{7}=\frac{4c_{2}-3c_{1}}{24\sqrt{3}}.
\end{align}
Consequently we find that the initial action \eqref{SUGRA} can be expressed in terms of powers of Weyl tensor
\begin{align}
	I_{5D,N=1,{\rm Weyl}}	&=\frac{\sigma}{16\pi}\int\sqrt{-g}d^{5}x(\mathcal{L}_{2\partial}+c_{1}\widetilde{\mathcal{L}}_{\text{Weyl}^{2}})\ , \label{SugraWeyl} \\
\mathcal{L}_{2\partial}	&=R+12\ell^{-2}-\frac{1}{4\tilde{g}^{2}}F^{\mu\nu}F_{\mu\nu}+\frac{\epsilon^{\mu\nu\rho\sigma\delta}F_{\mu\nu}F_{\rho\sigma}A_{\delta}}{12\sqrt{3}\tilde{g}^{3}}\ ,\nonumber\\
\widetilde{\mathcal{L}}_{\text{Weyl}^{2}}&=-\frac{2F^{\mu\nu}F_{\mu\nu}}{\tilde{g}^{2}\ell^{2}}-\frac{\epsilon^{\mu\nu\rho\sigma\delta}F_{\mu\nu}F_{\rho\sigma}A_{\delta}}{\sqrt{3}\tilde{g}^{3}\ell^{2}}+C_{\mu\nu\rho\sigma}C^{\mu\nu\rho\sigma}-\frac{C_{\mu\nu\rho\sigma}F^{\mu\nu}F^{\rho\sigma}}{2\tilde{g}^{2}}\nonumber\\
&+\frac{13(F^{\mu\nu}F_{\mu\nu})^{2}}{96\tilde{g}^{4}}-\frac{13}{24\tilde{g}^{4}}F_{\mu\nu}F^{\nu\lambda}F_{\lambda\delta}F^{\delta\mu}+\frac{\epsilon^{\mu\nu\rho\sigma\alpha}A_{\mu}C_{\nu\rho}^{\ \ \beta\gamma}C_{\sigma\alpha\beta\gamma}}{2\sqrt{3}\tilde{g}}\ ,\nonumber
\end{align}
where
\begin{align}
{\sigma}&=\sigma_{0}\big(1-24(c_{2}+c_{3})\ell_{0}^{-2}\big),\ \ {\tilde{g}}=\tilde{g}_{0}\big(1+4(c_{2}+c_{3})\ell_{0}^{-2}\big),\nonumber\\
{\ell}&=\ell_{0}\big(1-4(c_{2}+c_{3})\ell_{0}^{-2}\big)\ ,
	\label{EffParameter}
\end{align}
Unlike the ungauged supergravity, in the gauged case, the curvature squared supergravity actions also contain 2-derivative terms.  It is worth mentioning that $\ell_{0}$, $\tilde{g}_{0}$ and $\sigma_{0}$
represent the bare physical parameters before the field redefinitions,
$\ell$, $\tilde{g}$ and $\sigma$ denote the effective parameters
following the field redefinitions, they encode contributions
from the original 4-derivative supergravity
action (\ref{SUGRA}), such as $\ell_{0},\ \tilde{g}_{0},\ \sigma_{0}$ and
$c_{2},\ c_{3}$. Unlike the combinations of curvature squared terms chosen in \cite{Bobev:2022bjm, Cassani:2022lrk}, here the Weyl tensor squared  does not renormalize the bare AdS radius, which allows one to directly apply the Reall-Santos method to obtain on-shell action of AdS black holes \cite{Hu:2023gru,Ma:2024ynp}.

\subsection{Holographic central charges}
Let us now consider the holographic central charges that can be extracted from the renormalized action \eqref{SugraWeyl} 
\begin{align}
I_{{\rm ren}}	&=I_{5D,N=1,{\rm Weyl}}+I_{{\rm surf}},  \\ 
I_{{\rm surf}}	&=\frac{\sigma}{16\pi}\int_{z=\epsilon}d^{4}x\sqrt{-h}\big[2K-(\frac{6}{\ell}+\frac{\ell}{2}\mathcal{R})+\log\frac{\epsilon^{2}}{\ell^{2}}{\cal A}_{4}+...\big],
\label{Isurf}
\end{align}
where ${\cal A}_{4}=\frac{\ell^{3}}{8}(\mathcal{R}^{ij}\mathcal{R}_{ij}-\frac{1}{3}\mathcal{R}^{2})+c_{1}\frac{\ell}{2}\mathcal{C}^{ijkl}\mathcal{C}_{ijkl}$. The notation ``...'' denotes the additional logarithmic counterterm proportional to $F_{ij}F^{ij}$ \cite{Taylor:2000xw}, which does not contribute to our case.  $K$ is the extrinsic curvature of the AdS boundary located at $z=\epsilon$ for $\epsilon\rightarrow 0$, and $\mathcal{R}$, ${\cal R}_{ij}$ and ${\cal C}_{ijkl}$ refer to the boundary Ricci scalar curvature, Ricci tensor and Weyl tensor, respectively. The logarithmic terms  induced by  Einstein gravity were well known \cite{Henningson:1998gx}. However, we also find that the bulk  Weyl tensor squared also induces a new logarithmic counterterm proportional to $c_1$. Notice that the new logarithmic counterterm proportional to $c_1$ is absent when the bulk spacetime dimension is even (see \cite{Grumiller:2013mxa} for $D=4$ case and \cite{Anastasiou:2020mik, Anastasiou:2023oro} for $D=6$ case). Moreover, all the logarithmic terms vanish for AdS black holes with $S^1\times M_3$ type boundary topology for $M_3$ being Einstein which is the case for rotating Kerr-AdS black hole. The coefficients of the logarithmic counterterms also imply the central charges in the dual CFT \cite{Fukuma:2001uf}
\begin{align}
    {\rm a}=\frac{\pi{\ell}^3}{8}{\sigma},\quad {\rm c}=\frac{\pi\ell^{3}}{8}\sigma(1+8c_{1}\ell^{-2})\ .
\label{ac}
\end{align}
Substituting the effective parameters $\sigma,\ell$ into \eqref{ac}, we find  the a,c charges are consistent with the  charges presented in \cite{Gold:2023ymc}. Note that  \eqref{RedefWeyl} and \cite{Gold:2023ymc}   perform different field redefinitions, the redefinition in the former preserves thermodynamic equivalence, whereas the one in the latter does not.  To obtain the action in the GB basis of 4-derivative invariants, we can perform the field redefinitions \eqref{RedefWeyl} with these  parameters
\begin{align}
\lambda_{0}&=\frac{4\left(5c_{1}+6(c_{2}+5c_{3})\right)}{9\ell_{0}^{2}}\ ,\ \lambda_{1}=\frac{8c_{1}}{3}\ ,\ \lambda_{2}=-\frac{c_{1}}{3}\ ,\nonumber\\
\lambda_{3}&=\frac{4c_{1}}{3}\ ,\ \lambda_{4}=-\frac{c_{1}}{4}\ ,\  \lambda_{5}=\lambda_{6}=\lambda_{7}=0\ ,
\end{align}
which transform  the  Weyl basis \eqref{SugraWeyl} to the GB basis \cite{Gold:2023ymc}.
\section{Quadratic curvature corrections to charged rotating black holes in 5D minimal gauged supergravity}
\label{Holography-section-2}
In this section,  we review an efficient way of computing the quadratic curvature corrections to thermodynamic quantities of rotating black holes in $5D$ minimal gauged supergravity \cite{Ma:2024ynp} extended by curvature squared invariants. The computation
of on-shell Euclidean action boils down to simply evaluating the uncorrected solution in the higher derivative action \cite{Hu:2023gru}, without being concerned about the complicated higher derivative surface
counterterms, which would be required in the ordinary approach. 
In particular, this approach is convenient for
studying rotating AdS black holes, because to obtain the higher
derivative corrections to the Euclidean action, one only
needs to evaluate a bulk integral which has no preference
on the choice of coordinates that may affect the induced metric on the conformal boundary \cite{Gibbons:2004ai}. Using this approach, we obtain quadratic curvature corrections to charged rotating black holes in $5D$ minimal gauged supergravity.

 In subsection \ref{Section3.1}, we have transformed the curvature-squared supergravity eq.\eqref{SUGRA} to the Weyl-squared supergravity eq. \eqref{SugraWeyl} which possesses equivalent thermodynamic variables as the original theory. The approach proposed in \cite{Hu:2023gru,Ma:2024ynp} suggests that the fully corrected on-shell action can be obtained by simply evaluating the modified action eq. \eqref{SugraWeyl} on the uncorrected solution, such as the general charged rotating black hole solution in 2-derivative $5D$ minimal gauged supergravity obtained in \cite{Chong:2005hr}.
 Regularity of the solution determines the inverse temperature to be
\begin{equation}
\beta=\frac{2\pi r_{0}\left(abq+(r_{0}^{2}+a^{2})(r_{0}^{2}+b^{2})\right)}{r_{0}^{4}\left[1+(a^{2}+b^{2}+2r_{0}^{2}){\ell}^{-2}\right]-(ab+q)^{2}}\ ,
\label{Tem}
\end{equation}
where $r_0$ is the radius of the outer horizon, $a,\,b,\,q$ are parameters related to angular velocities and electrostatic potential given by
\begin{eqnarray}    
&&{\Omega}_{a}=\frac{a(r_{0}^{2}+b^{2})(1+{\ell}^{-2}r_{0}^{2})+bq}{abq+(r_{0}^{2}+a^{2})(r_{0}^{2}+b^{2})}\ ,\nonumber\\
&&{\Omega}_{b}=\frac{b(r_{0}^{2}+a^{2})(1+{\ell}^{-2}r_{0}^{2})+aq}{abq+(r_{0}^{2}+a^{2})(r_{0}^{2}+b^{2})}\ , \nonumber\\
&&{\Phi}_{e}=\frac{\tilde{g}\sqrt{3}qr_{0}^{2}}{abq+(r_{0}^{2}+a^{2})(r_{0}^{2}+b^{2})}\ .
\label{tp}
\end{eqnarray}

 The on-shell Euclidean action should be viewed as a function of $T$, $\Omega_{a,b}$ and $\Phi_e$, as the current choice of boundary condition specifies the grand canonical ensemble.  Then we obtain the Euclidean action for the general charged rotating AdS black holes in $5D$ minimal gauged supergravity extended by all three curvature-squared invariants eq.\eqref{SugraWeyl}
\begin{align}
I_{\rm ren}(T,\Phi_e,\Omega_a,\Omega_b)=\frac{\pi\beta\sigma}{4\Xi_{a}\Xi_{b}}\big(m-\frac{q^{2}r_{0}^{2}}{X+abq}-X\ell^{-2}\big)-\frac{c_{1}\pi\beta\sigma\Sigma_{k=0}^{8}d_{2k}r_{0}^{2k}}{4r_{0}^{4}\ell^{4}\Xi_{a}\Xi_{b}X\big(X+abq\big)}\ , 
\label{FullAction}
\end{align}
where $m=2r_{0}^{-2}\big(2abq+q^{2}+X(1+r_{0}^{2}\ell^{-2})\big),\ X=(a^{2}+r_{0}^{2})(b^{2}+r_{0}^{2}),\ \Xi_{a}=1-a^2\ell^{-2},\ \Xi_{b}=1-b^2\ell^{-2},\  y=\frac{q}{ab},\ Y_{m,n}=(ab)^{m}(a^{n}+b^{n})$. 
The effects of the supersymmetric Ricci tensor squared and the Ricci scalar squared are encoded in the effective AdS radius. The coefficients $d_{2k}$ are 
\begin{align}
d_{0}&=\frac{1}{2}Y_{6,0}\ell^{4}(y+1)^{5},\ \ \ d_{2}=\ell^{2}(y+1)^{3}(Y_{6,0}-\ell^{2}(3y+5)Y_{4,2})\ ,\nonumber\\
d_{4}&=\frac{1}{2}(y+1)\big[-\ell^{4}\big((3y^{3}+19y^{2}+45y+31)Y_{4,0}+2(2y+5)Y_{2,4}\big)\nonumber\\
&+Y_{6,0}-4(3y^{2}+9y+5)\ell^{2}Y_{4,2}\big]\ ,\nonumber\\
d_{6}&=\ell^{4}\big(Y_{0,6}-(8y^{2}+26y+23)Y_{2,2}\big)-5(2y+1)Y_{4,2}\nonumber\\
&-2\ell^{2}\big[(5-y^{2}+y)Y_{2,4}+\frac{1}{2}\big(31-y(y-26)(y+2)\big)Y_{4,0}\big]\ ,\nonumber\\
d_{8}&=\ell^{4}\big(3Y_{0,4}-\frac{1}{2}(6y^{2}+3y+7)Y_{2,0}\big)+2\ell^{2}\big((4y^{2}-6y-23)Y_{2,2}+Y_{0,6}\big)\nonumber\\
&+5(y-1)Y_{2,4}-\frac{1}{2}(28y+31)Y_{4,0}\ ,\nonumber\\
d_{10}&=11\ell^{4}Y_{0,2}+\ell^{2}\big((3y^{2}+11y-7)Y_{2,0}+6Y_{0,4}\big)+(14y-23)Y_{2,2}+Y_{0,6}\ ,\nonumber\\
d_{12}&=9\ell^{4}+22\ell^{2}Y_{0,2}+\frac{1}{2}(25y-7)Y_{2,0}+3Y_{0,4}\ ,\nonumber\\
d_{14}&=18\ell^{2}+11Y_{0,2}, d_{16}=9\ .
\end{align}

The other thermodynamic
variables are obtained from the action \eqref{FullAction} via standard relations
\begin{eqnarray}
  M&=&G+TS+\Omega_{a}J_{a}+\Omega_{b}J_{b},\quad S=-\frac{\partial G}{\partial T}|_{\Phi_{e},\Omega_{a,b}},
  \nonumber\\
  \ Q_{e}&=&-\frac{\partial G}{\partial\Phi_{e}}|_{T,\Omega_{a,b}},\ J_{a(b)}=-\frac{\partial G}{\partial\Omega_{a(b)}}|_{T,\Phi_{e},\Omega_{b(a)}}\ ,
\end{eqnarray}
where  $G=TI_{\text{ren}}$ denotes the Gibbs free energy. We now apply the result to understand the entropy of the supersymmetric charged rotating AdS black hole \cite{Gutowski:2004ez,Gutowski:2004yv} which admits a microscopic description in terms of the index of the dual $4D,\,{\mathcal{N}}=1$ superconformal field theory \cite{Cabo-Bizet:2018ehj, Choi:2018hmj,Benini:2018ywd, Honda:2019cio, David:2020ems, Agarwal:2020zwm, Benini:2020gjh}. To proceed, we impose the supersymmetry condition
	\begin{equation}
	q=-(a-ir_{0})(b-ir_{0})(1-ir_{0}{\ell}^{-1})\ .
	\label{susy}
	\end{equation}
	Note that the BPS limit also requires zero temperature and can be reached via $r_{0}\rightarrow\sqrt{{\ell}(a+b)+ab}$ \cite{Chong:2005hr}. Unlike previous works \cite{Bobev:2022bjm,Cassani:2022lrk}, the supersymmetric condition is now corrected by the 4-derivative terms whose effect is fully encoded in the effective AdS radius $\ell$. Subsequently, one can define thermodynamic potentials
	\begin{align}
	\omega_{a}&={\beta}_{s}({\Omega}_{a,s}-{\Omega}_{a,*})=\frac{2\pi(b-ir_{0})(a-{\ell})}{\Xi}\ ,\nonumber \\
	\omega_{b}&={\beta}_{s}({\Omega}_{b,s}-{\Omega}_{b,*})=\frac{2\pi(a-ir_{0})(b-{\ell})}{\Xi}\ ,\nonumber \\
	\varphi &={\beta}_{s}({\Phi}_{e,s}-{\Phi}_{e,*})=\frac{6{\pi \tilde{g}\ell  }(a-ir_{0})(b-ir_{0})}{\sqrt{3}\Xi},
    \label{oabf}
	\end{align}
	where $\Xi= 2r_{0}({\ell}+a+b)+i\big({\ell}(a+b)+ab\big)-3ir_{0}^{2}$ and they
satisfy $
	\omega_{a}+\omega_{b}-\frac{\sqrt{3}}{\tilde{g}{\ell}}\varphi-2\pi i=0$.
	Here ``$s$" means the supersymmetry condition \eqref{susy} has been applied and ``*" denotes the values of these variable in the BPS limit
	\begin{equation}
	{\Omega}_{a,*}={\ell}^{-1}\ ,\quad {\Omega}_{b,*}={\ell}^{-1}\ ,\quad {\Phi}_{e,*}=\sqrt{3}\tilde{g}\ .
	\end{equation}
	Imposing the supersymmetric condition \eqref{susy}, we find that the Euclidean action drastically simplifies
	\begin{equation}
	I_{{\rm ren},s}=\frac{\pi\sigma\varphi^{3}(1-\frac{12c_{3}}{\ell^{2}})}{12\sqrt{3}\tilde{g}^{3}\omega_{a}\omega_{b}}+\frac{c_{1}\pi\sigma\varphi(\omega_{a}^{2}+\omega_{b}^{2}-4\pi^{2})}{\sqrt{3}\tilde{g}\omega_{a}\omega_{b}}\ .
	\label{regI}
	\end{equation}
It is worth approaching this result from a different perspective. We can consider the following action
   \begin{align}
	I&=\frac{\sigma}{16\pi}\int\sqrt{-g}d^{5}x(\mathcal{L}_{2\partial}+c_{1}\widetilde{\mathcal{L}}_{\text{Weyl}^{2}})\ , \nonumber \\
\widetilde{\mathcal{L}}_{\text{Weyl}^{2}}	&=c'_{1}\frac{F^{\mu\nu}F_{\mu\nu}}{\tilde{g}^{2}\ell^{2}}+c'_{2}\frac{\epsilon^{\mu\nu\rho\sigma\delta}F_{\mu\nu}F_{\rho\sigma}A_{\delta}}{\tilde{g}^{3}\ell^{2}}+c'_{3}C_{\mu\nu\rho\sigma}C^{\mu\nu\rho\sigma}+c'_{4}\frac{C_{\mu\nu\rho\sigma}F^{\mu\nu}F^{\rho\sigma}}{\tilde{g}^{2}}\nonumber\\
&+c'_{5}\frac{(F^{\mu\nu}F_{\mu\nu})^{2}}{\tilde{g}^{4}}+\frac{c'_{6}}{\tilde{g}^{4}}F_{\mu\nu}F^{\nu\lambda}F_{\lambda\delta}F^{\delta\mu}+\frac{c'_{7}}{\tilde{g}}\epsilon^{\mu\nu\rho\sigma\alpha}A_{\mu}C_{\nu\rho}^{\ \ \beta\gamma}C_{\sigma\alpha\beta\gamma}\ .  \label{SugraWeyl2}
\end{align}
We find that requiring the action \eqref{SugraWeyl2} with the supersymmetric condition to depend solely on $\omega_{a,b}$ uniquely determines the coefficients, so that the resulting action recovers the supergravity combination \eqref{SugraWeyl}. This offers an indirect method to deduce the supergravity combination. Although our supersymmetric action takes the same form as those results in  \cite{Bobev:2022bjm,Cassani:2022lrk}, the details are different, as our $\omega_{a,b},\varphi$ are defined in \eqref{oabf} and depend on the effective AdS radius, while those in \cite{Bobev:2022bjm,Cassani:2022lrk} used bare AdS radius instead. 
Taking the BPS limit, we find that the conserved charges obey the linear relation \footnote{A similar equality was obtained in \cite{Cassani:2022lrk}, where it is the bare AdS radius rather than the effective AdS radius that enters the expression. }
	\begin{equation}
	M_{*}-{\ell}^{-1}J_{a,*}-{\ell}^{-1}J_{b,*}-\frac{3}{2}{\ell}^{-1}Q_{R}=0\ ,
	\end{equation}
	where $Q_{R}:=\frac{2\tilde{g}{\ell}}{\sqrt{3}}Q_{e,*}$ is the canonically normalized $U(1)_R$-charge in the dual SCFT. This equality leads to vanishing Gibbs free energy. In the BPS limit, the entropy of the charged rotating black hole also reproduces the microscopic result \cite{Bobev:2022bjm, Cassani:2022lrk}. Namely, up to ${\cal O}(c_i)$ it's given by
		\begin{equation}
		S_{*}=\pi\sqrt{3Q_{R}^{2}-8\mathrm{a}\left(J_{a,*}+J_{b,*}\right)
			-16\mathrm{a}(\mathrm{a}-\mathrm{c})\frac{\left(J_{a,*}
				-J_{b,*}\right){}^{2}}{Q_{R}^{2}-2\mathrm{a}\left(J_{a,*}+J_{b,*}\right)}}\ .
		\end{equation}
One can verify that these BPS charges  satisfy non-linear relations
\begin{eqnarray}
   && \big(3Q_{R}+4(2\mathrm{a}-\mathrm{c})\big)\big(3Q_{R}^{2}-8\mathrm{c}(J_{a,*}+J_{b,*})\big)\nonumber \\
&=& Q_{R}^{3}+16(3\mathrm{c}-2\mathrm{a})J_{a,*}J_{b,*}+64\mathrm{a}(\mathrm{a}-\mathrm{c})\frac{(Q_{R}+\mathrm{a})(J_{a,*}-J_{b,*})^{2}}{Q_{R}^{2}-2\mathrm{a}(J_{a,*}+J_{b,*})}\ .
\label{non-linear} 
\end{eqnarray}

\section{Conclusion and Outlook}
\label{Conclusion-section}

In this work, we reviewed some of our recent developments in constructing curvature-squared invariants in five-dimensional minimal gauged supergravity. Using superspace and superconformal tensor calculus in a framework based on the gauged dilaton-Weyl multiplet, we outlined the construction of the supersymmetric completions of curvature-squared terms, including the notably intricate Ricci tensor squared (Log multiplet). These advances significantly enhance our capability to study quantum aspects of gravity and holography within a controlled supersymmetric setting.

An achievement of this analysis was the simplification of on-shell expressions via field redefinitions, which allowed for clear identification of holographic central charges and facilitated direct comparison with AdS/CFT predictions. We derived explicit on-shell expressions and showed consistency with expected dual-field theory results stressing simplifications based on using the Weyl-squared combination as the surviving $\alpha^\prime$ correction after going on-shell. Additionally, we introduced an efficient framework to compute higher-curvature corrections to the thermodynamics of charged, rotating AdS$_5$ black holes, verifying known supersymmetric limits and uncovering novel linear and nonlinear relations among conserved quantities.

Several promising directions remain open for future investigation.
Extending this framework to include additional matter multiplets and theories with extended supersymmetry, especially in contexts motivated by string theory, would significantly broaden its applicability.
While our focus was on curvature-squared terms, generalizing the construction to incorporate higher derivative invariants may reveal new structures in quantum gravity and refine holographic precision tests.
In particular, one might wonder whether a special role is played by four-derivative terms and their Chern-Simons couplings. It would be remarkable if, in five dimensions, 
higher-order, CS-free terms, would not contribute to BPS observables, both in the flat and AdS case. We hope to report on first promising results in this direction in the near future \cite{HKPT-M_progress}. 
Matching higher-derivative-corrected asymptotically ${\rm AdS}_5$ black hole entropy with subleading corrections from dual CFT microstate counts would provide a stringent test of holographic duality and is part of a currently active field of research, see, e.g., \cite{Bobev:2022bjm,Cassani:2022lrk,Ma:2024ynp,Cassani:2024tvk,Cassani:2025sim} for recent works and review on this direction. Our results and further higher derivative extension would play an important role in this area.
A powerful tool that has recently  been  employed in supergravity with applications to holography is equivariant localization --- see  \cite{Hristov:2024cgj} for a pedagogical review and
\cite{BenettiGenolini:2023ndb, BenettiGenolini:2023kxp,Cassani:2024kjn,Colombo:2025ihp} for recent results. 
A very interesting open question is to extend these analyses to the case of (off-shell) supergravity with higher-derivatives corrections.
More in general, further development of holographic renormalization techniques for higher-derivative supergravity actions, especially in nontrivial boundary geometries or dimensions, remains an important open task --- see \cite{Cassani:2023vsa} for some recent analyses in that direction.
Constructing and analyzing full black hole solutions, either numerically or analytically, in the presence of higher-derivative corrections would help test the general framework and potentially uncover new physics.

We hope that the methods and results presented here will serve as a useful reference and a springboard for further work on higher-derivative supergravities and their applications in precision tests of holography.

\begin{acknowledgement}
We are grateful to K.\,Hristov, S.\,M.\,Kuzenko, E.\,Raptakis, C.\,Kennedy and to all the other participants of the MATRIX program ``New Deformations of Quantum Field and Gravity Theories'' for stimulating discussions and collaborations related to the subject of this work.
P. Hu and Y. Pang are supported by the National Key R\&D Program No. 2022YFE0134300 and
the National Natural Science Foundation of China (NSFC) Grant No. 12175164.
The work of G.T.-M. was supported by the Australian Research Council (ARC) Future Fellowship FT180100353, ARC Discovery Project DP240101409, and the Capacity Building Package of the University of Queensland.
G.G. and S.K. were supported by the postgraduate scholarships at the University of Queensland.
The work of J.H. is also supported by the European Union under the Marie Sklodowska-Curie grant agreement number 101107602.\footnote{Views and opinions expressed are however those of the author(s) only and do not necessarily reflect those of the European Union or European Research Executive Agency. Neither the European Union nor the granting authority can be held responsible for them.}
\end{acknowledgement}
%


\bigskip
%

\begin{thebibliography}{99.}%
%
%
%

\providecommand{\url}[1]{{#1}}
\providecommand{\urlprefix}{URL }
\expandafter\ifx\csname urlstyle\endcsname\relax
  \providecommand{\doi}[1]{DOI~\discretionary{}{}{}#1}\else
  \providecommand{\doi}{DOI~\discretionary{}{}{}\begingroup \urlstyle{rm}\Url}\fi


\bibitem{Agarwal:2020zwm}
Agarwal, P., Choi, S., Kim, J., Kim, S., Nahmgoong, J.: {AdS black holes and finite N indices}.
\newblock Phys. Rev. D \textbf{103}(12), 126,006 (2021).
\newblock \doi{10.1103/PhysRevD.103.126006}

\bibitem{Anastasiou:2023oro}
Anastasiou, G., Araya, I.J., Corral, C., Olea, R.: {Conformal Renormalization of topological black holes in AdS$_{6}$}.
\newblock JHEP \textbf{11}, 036 (2023).
\newblock \doi{10.1007/JHEP11(2023)036}

\bibitem{Anastasiou:2020mik}
Anastasiou, G., Araya, I.J., Olea, R.: {Einstein Gravity from Conformal Gravity in 6D}.
\newblock JHEP \textbf{01}, 134 (2021).
\newblock \doi{10.1007/JHEP01(2021)134}

\bibitem{BenettiGenolini:2023kxp}
Benetti~Genolini, P., Gauntlett, J.P., Sparks, J.: {Equivariant Localization in Supergravity}.
\newblock Phys. Rev. Lett. \textbf{131}(12), 121,602 (2023).
\newblock \doi{10.1103/PhysRevLett.131.121602}

\bibitem{BenettiGenolini:2023ndb}
Benetti~Genolini, P., Gauntlett, J.P., Sparks, J.: {Equivariant localization for AdS/CFT}.
\newblock JHEP \textbf{02}, 015 (2024).
\newblock \doi{10.1007/JHEP02(2024)015}

\bibitem{Benini:2020gjh}
Benini, F., Colombo, E., Soltani, S., Zaffaroni, A., Zhang, Z.: {Superconformal indices at large $N$ and the entropy of AdS$_5$ $\times$ SE$_5$ black holes}.
\newblock Class. Quant. Grav. \textbf{37}(21), 215,021 (2020).
\newblock \doi{10.1088/1361-6382/abb39b}

\bibitem{Benini:2018ywd}
Benini, F., Milan, E.: {Black Holes in 4D $\mathcal{N}$=4 Super-Yang-Mills Field Theory}.
\newblock Phys. Rev. X \textbf{10}(2), 021,037 (2020).
\newblock \doi{10.1103/PhysRevX.10.021037}

\bibitem{Bergshoeff:2002qk}
Bergshoeff, E., Cucu, S., De~Wit, T., Gheerardyn, J., Halbersma, R., Vandoren, S., Van~Proeyen, A.: {Superconformal N=2, D = 5 matter with and without actions}.
\newblock JHEP \textbf{10}, 045 (2002).
\newblock \doi{10.1088/1126-6708/2002/10/045}

\bibitem{Bergshoeff:2004kh}
Bergshoeff, E., Cucu, S., de~Wit, T., Gheerardyn, J., Vandoren, S., Van~Proeyen, A.: {N = 2 supergravity in five-dimensions revisited}.
\newblock Class. Quant. Grav. \textbf{21}, 3015--3042 (2004).
\newblock \doi{10.1088/0264-9381/23/23/C01}

\bibitem{Bergshoeff:2001hc}
Bergshoeff, E., de~Wit, T., Halbersma, R., Cucu, S., Derix, M., Van~Proeyen, A.: {Weyl multiplets of N=2 conformal supergravity in five-dimensions}.
\newblock JHEP \textbf{06}, 051 (2001).
\newblock \doi{10.1088/1126-6708/2001/06/051}

\bibitem{Bergshoeff:2011xn}
Bergshoeff, E.A., Rosseel, J., Sezgin, E.: {Off-shell D=5, N=2 Riemann Squared Supergravity}.
\newblock Class. Quant. Grav. \textbf{28}, 225,016 (2011).
\newblock \doi{10.1088/0264-9381/28/22/225016}

\bibitem{Bobev:2022bjm}
Bobev, N., Dimitrov, V., Reys, V., Vekemans, A.: {Higher derivative corrections and AdS5 black holes}.
\newblock Phys. Rev. D \textbf{106}(12), L121,903 (2022).
\newblock \doi{10.1103/PhysRevD.106.L121903}

\bibitem{Butter:2014xxa}
Butter, D., Kuzenko, S.M., Novak, J., Tartaglino-Mazzucchelli, G.: {Conformal supergravity in five dimensions: New approach and applications}.
\newblock JHEP \textbf{02}, 111 (2015).
\newblock \doi{10.1007/JHEP02(2015)111}

\bibitem{Cabo-Bizet:2018ehj}
Cabo-Bizet, A., Cassani, D., Martelli, D., Murthy, S.: {Microscopic origin of the Bekenstein-Hawking entropy of supersymmetric AdS$_{5}$ black holes}.
\newblock JHEP \textbf{10}, 062 (2019).
\newblock \doi{10.1007/JHEP10(2019)062}

\bibitem{Cassani:2025sim}
Cassani, D., Murthy, S.: {Quantum black holes: supersymmetry and exact results} (2025)

\bibitem{Cassani:2022lrk}
Cassani, D., Ruip\'erez, A., Turetta, E.: {Corrections to AdS$_{5}$ black hole thermodynamics from higher-derivative supergravity}.
\newblock JHEP \textbf{11}, 059 (2022).
\newblock \doi{10.1007/JHEP11(2022)059}

\bibitem{Cassani:2023vsa}
Cassani, D., Ruip\'erez, A., Turetta, E.: {Boundary terms and conserved charges in higher-derivative gauged supergravity}.
\newblock JHEP \textbf{06}, 203 (2023).
\newblock \doi{10.1007/JHEP06(2023)203}

\bibitem{Cassani:2024tvk}
Cassani, D., Ruip\'erez, A., Turetta, E.: {Higher-derivative corrections to flavoured BPS black hole thermodynamics and holography}.
\newblock JHEP \textbf{05}, 276 (2024).
\newblock \doi{10.1007/JHEP05(2024)276}

\bibitem{Cassani:2024kjn}
Cassani, D., Ruip\'erez, A., Turetta, E.: {Localization of the 5D supergravity action and Euclidean saddles for the black hole index}.
\newblock JHEP \textbf{12}, 086 (2024).
\newblock \doi{10.1007/JHEP12(2024)086}

\bibitem{CN}
Chamseddine, A.H., Nicolai, H.: {Coupling the SO(2) Supergravity Through Dimensional Reduction}.
\newblock Phys. Lett. B \textbf{96}, 89--93 (1980).
\newblock \doi{10.1016/0370-2693(80)90218-X}.
\newblock [Erratum: Phys.Lett.B 785, 631--632 (2018)]

\bibitem{Choi:2018hmj}
Choi, S., Kim, J., Kim, S., Nahmgoong, J.: {Large AdS black holes from QFT}  (2018)

\bibitem{Chong:2005hr}
Chong, Z.W., Cvetic, M., Lu, H., Pope, C.N.: {General non-extremal rotating black holes in minimal five-dimensional gauged supergravity}.
\newblock Phys. Rev. Lett. \textbf{95}, 161,301 (2005).
\newblock \doi{10.1103/PhysRevLett.95.161301}

\bibitem{Colombo:2025ihp}
Colombo, E., Dimitrov, V., Martelli, D., Zaffaroni, A.: {Equivariant localization in supergravity in odd dimensions}  (2025)

\bibitem{Coomans:2012cf}
Coomans, F., Ozkan, M.: {An off-shell formulation for internally gauged D=5, N=2 supergravity from superconformal methods}.
\newblock JHEP \textbf{01}, 099 (2013).
\newblock \doi{10.1007/JHEP01(2013)099}

\bibitem{Cremmer}
Cremmer, E.: {Supergravities in 5 Dimensions}.
\newblock In: {Supergravity and Superspace, S. W. Hawking and M. Rocek (Eds.)}, pp. 267--282. Cambridge Univ. Press, Cambridge, UK (1980)

\bibitem{David:2020ems}
David, M., Nian, J., Pando~Zayas, L.A.: {Gravitational Cardy Limit and AdS Black Hole Entropy}.
\newblock JHEP \textbf{11}, 041 (2020).
\newblock \doi{10.1007/JHEP11(2020)041}

\bibitem{Deser:1986xr}
Deser, S., Redlich, A.N.: {String Induced Gravity and Ghost Freedom}.
\newblock Phys. Lett. B \textbf{176}, 350 (1986).
\newblock \doi{10.1016/0370-2693(86)90177-2}.
\newblock [Erratum: Phys.Lett.B 186, 461 (1987)]

\bibitem{Freedman:2012zz}
Freedman, D.Z., Van~Proeyen, A.: {Supergravity}.
\newblock Cambridge Univ. Press, Cambridge, UK (2012).
\newblock \doi{10.1017/CBO9781139026833}

\bibitem{Fujita:2001kv}
Fujita, T., Ohashi, K.: {Superconformal tensor calculus in five-dimensions}.
\newblock Prog. Theor. Phys. \textbf{106}, 221--247 (2001).
\newblock \doi{10.1143/PTP.106.221}

\bibitem{Fukuma:2001uf}
Fukuma, M., Matsuura, S., Sakai, T.: {Higher derivative gravity and the AdS / CFT correspondence}.
\newblock Prog. Theor. Phys. \textbf{105}, 1017--1044 (2001).
\newblock \doi{10.1143/PTP.105.1017}

\bibitem{Gibbons:2004ai}
Gibbons, G.W., Perry, M.J., Pope, C.N.: {The First law of thermodynamics for Kerr-anti-de Sitter black holes}.
\newblock Class. Quant. Grav. \textbf{22}, 1503--1526 (2005).
\newblock \doi{10.1088/0264-9381/22/9/002}

\bibitem{Gold:2023ymc}
Gold, G., Hutomo, J., Khandelwal, S., Ozkan, M., Pang, Y., Tartaglino-Mazzucchelli, G.: {All Gauged Curvature-Squared Supergravities in Five Dimensions}.
\newblock Phys. Rev. Lett. \textbf{131}(25), 251,603 (2023).
\newblock \doi{10.1103/PhysRevLett.131.251603}

\bibitem{Gold:2023dfe}
Gold, G., Hutomo, J., Khandelwal, S., Tartaglino-Mazzucchelli, G.: {Curvature-squared invariants of minimal five-dimensional supergravity from superspace}.
\newblock Phys. Rev. D \textbf{107}(10), 106,013 (2023).
\newblock \doi{10.1103/PhysRevD.107.106013}

\bibitem{Gold:2023ykx}
Gold, G., Hutomo, J., Khandelwal, S., Tartaglino-Mazzucchelli, G.: {Components of curvature-squared invariants of minimal supergravity in five dimensions}.
\newblock JHEP \textbf{07}, 221 (2024).
\newblock \doi{10.1007/JHEP07(2024)221}

\bibitem{Gold:2024nbw}
Gold, G., Khandelwal, S., Tartaglino-Mazzucchelli, G.: {Supergravity Component Reduction with Computer Algebra} (2024)

\bibitem{Grumiller:2013mxa}
Grumiller, D., Irakleidou, M., Lovrekovic, I., McNees, R.: {Conformal gravity holography in four dimensions}.
\newblock Phys. Rev. Lett. \textbf{112}, 111,102 (2014).
\newblock \doi{10.1103/PhysRevLett.112.111102}

\bibitem{Gutowski:2004yv}
Gutowski, J.B., Reall, H.S.: {General supersymmetric AdS(5) black holes}.
\newblock JHEP \textbf{04}, 048 (2004).
\newblock \doi{10.1088/1126-6708/2004/04/048}

\bibitem{Gutowski:2004ez}
Gutowski, J.B., Reall, H.S.: {Supersymmetric AdS(5) black holes}.
\newblock JHEP \textbf{02}, 006 (2004).
\newblock \doi{10.1088/1126-6708/2004/02/006}

\bibitem{HOT}
Hanaki, K., Ohashi, K., Tachikawa, Y.: {Supersymmetric Completion of an $R^2$ term in Five-dimensional Supergravity}.
\newblock Prog. Theor. Phys. \textbf{117}, 533 (2007).
\newblock \doi{10.1143/PTP.117.533}

\bibitem{Henningson:1998gx}
Henningson, M., Skenderis, K.: {The Holographic Weyl anomaly}.
\newblock JHEP \textbf{07}, 023 (1998).
\newblock \doi{10.1088/1126-6708/1998/07/023}

\bibitem{Honda:2019cio}
Honda, M.: {Quantum Black Hole Entropy from 4d Supersymmetric Cardy formula}.
\newblock Phys. Rev. D \textbf{100}(2), 026,008 (2019).
\newblock \doi{10.1103/PhysRevD.100.026008}

\bibitem{Howe5Dsugra}
Howe, P.S.: {Off-shell N=2 and N=4 supergravity in five-dimensions}.
\newblock In: {Nuffield Workshop on Quantum Structure of Space and Time, M. J. Duff and C. J. Isham (Eds.)}, pp. 239--253. Cambridge Univ. Press, Cambridge, UK (1981)

\bibitem{Hristov:2024cgj}
Hristov, K.: {Equivariant localization and gluing rules in 4d $\mathcal{N}=2$ higher derivative supergravity} (2024)

\bibitem{Hu:2023gru}
Hu, P.J., Ma, L., L\"u, H., Pang, Y.: {Improved Reall-Santos method for AdS black holes in general 4-derivative gravities}.
\newblock Sci. China Phys. Mech. Astron. \textbf{67}(8), 280,412 (2024).
\newblock \doi{10.1007/s11433-024-2398-1}

\bibitem{Hutomo:2022hdi}
Hutomo, J., Khandelwal, S., Tartaglino-Mazzucchelli, G., Woods, J.: {Hyperdilaton Weyl multiplets of 5D and 6D minimal conformal supergravity}.
\newblock Phys. Rev. D \textbf{107}(4), 046,009 (2023).
\newblock \doi{10.1103/PhysRevD.107.046009}

\bibitem{HKPT-M_progress}
K.~Hristov S.~Khandelwal, Y.P., Tartaglino-Mazzucchelli, G.: Work in progress

\bibitem{Kugo:2000af}
Kugo, T., Ohashi, K.: {Off-shell D = 5 supergravity coupled to matter Yang-Mills system}.
\newblock Prog. Theor. Phys. \textbf{105}, 323--353 (2001).
\newblock \doi{10.1143/PTP.105.323}

\bibitem{Kugo:2002vc}
Kugo, T., Ohashi, K.: {Gauge and nongauge tensor multiplets in 5-D conformal supergravity}.
\newblock Prog. Theor. Phys. \textbf{108}, 1143--1164 (2003).
\newblock \doi{10.1143/PTP.108.1143}

\bibitem{Kuzenko:2022ajd}
Kuzenko, S.M., Raptakis, E.S.N., Tartaglino-Mazzucchelli, G.: {Covariant Superspace Approaches to  N=2 Supergravity} (2023).
\newblock \doi{10.1007/978-981-19-3079-9\_44-1}

\bibitem{Kuzenko:2022skv}
Kuzenko, S.M., Raptakis, E.S.N., Tartaglino-Mazzucchelli, G.: {Superspace Approaches to N} = 1 Supergravity (2023).
\newblock \doi{10.1007/978-981-19-3079-9\_40-1}

\bibitem{Lauria:2020rhc}
Lauria, E., Van~Proeyen, A.: {${\cal N}=2$ Supergravity in $D=4,5,6$ Dimensions}, vol. 966 (2020).
\newblock \doi{10.1007/978-3-030-33757-5}

\bibitem{Liu:2022sew}
Liu, J.T., Saskowski, R.J.: {Four-derivative corrections to minimal gauged supergravity in five dimensions}.
\newblock JHEP \textbf{05}, 171 (2022).
\newblock \doi{10.1007/JHEP05(2022)171}

\bibitem{Ma:2024ynp}
Ma, L., Hu, P.J., Pang, Y., Lu, H.: {Effectiveness of Weyl gravity in probing quantum corrections to AdS black holes}.
\newblock Phys. Rev. D \textbf{110}(2), L021,901 (2024).
\newblock \doi{10.1103/PhysRevD.110.L021901}

\bibitem{Metsaev:1987zx}
Metsaev, R.R., Tseytlin, A.A.: {Order alpha-prime (Two Loop) Equivalence of the String Equations of Motion and the Sigma Model Weyl Invariance Conditions: Dependence on the Dilaton and the Antisymmetric Tensor}.
\newblock Nucl. Phys. B \textbf{293}, 385--419 (1987).
\newblock \doi{10.1016/0550-3213(87)90077-0}

\bibitem{Ozkan:2013nwa}
Ozkan, M., Pang, Y.: {All off-shell $R^{2}$ invariants in five dimensional $\mathcal{N} =$ 2 supergravity}.
\newblock JHEP \textbf{08}, 042 (2013).
\newblock \doi{10.1007/JHEP08(2013)042}

\bibitem{Ozkan:2013uk}
Ozkan, M., Pang, Y.: {Supersymmetric Completion of Gauss-Bonnet Combination in Five Dimensions}.
\newblock JHEP \textbf{03}, 158 (2013).
\newblock \doi{10.1007/JHEP07(2013)152}.
\newblock [Erratum: JHEP 07, 152 (2013)]

\bibitem{Ozkan:2024euj}
Ozkan, M., Pang, Y., Sezgin, E.: {Higher derivative supergravities in diverse dimensions}.
\newblock Phys. Rept. \textbf{1086}, 1--95 (2024).
\newblock \doi{10.1016/j.physrep.2024.07.002}

\bibitem{Cadabra-1}
Peeters, K.: {A Field-theory motivated approach to symbolic computer algebra}.
\newblock Comput. Phys. Commun. \textbf{176}, 550--558 (2007).
\newblock \doi{10.1016/j.cpc.2007.01.003}

\bibitem{Cadabra-2}
Peeters, K.: {Introducing Cadabra: A Symbolic computer algebra system for field theory problems}  (2007)

\bibitem{Reall:2019sah}
Reall, H.S., Santos, J.E.: {Higher derivative corrections to Kerr black hole thermodynamics}.
\newblock JHEP \textbf{04}, 021 (2019).
\newblock \doi{10.1007/JHEP04(2019)021}

\bibitem{Taylor:2000xw}
Taylor, M.: {More on counterterms in the gravitational action and anomalies}  (2000)

\bibitem{Zumino:1985dp}
Zumino, B.: {Gravity Theories in More Than Four-Dimensions}.
\newblock Phys. Rept. \textbf{137}, 109 (1986).
\newblock \doi{10.1016/0370-1573(86)90076-1}

\bibitem{Zwiebach:1985uq}
Zwiebach, B.: {Curvature Squared Terms and String Theories}.
\newblock Phys. Lett. B \textbf{156}, 315--317 (1985).
\newblock \doi{10.1016/0370-2693(85)91616-8}



\end{thebibliography}
%

\end{document}